\documentclass[10pt,twocolumn,letterpaper]{article}

\usepackage[pagenumbers]{cvpr} 
\usepackage{times}
\usepackage{epsfig}
\usepackage{graphicx}
\usepackage{amsmath}
\usepackage{amssymb}
\usepackage{booktabs}
\usepackage{float}
\usepackage{array}
\usepackage{multirow}
\usepackage[table]{xcolor}
\usepackage{enumitem}
\usepackage{tikz}
\usetikzlibrary{arrows.meta, positioning, calc, shapes.geometric, decorations.pathreplacing}

\definecolor{cGreen}{HTML}{2E7D32}
\definecolor{cRed}{HTML}{C62828}
\definecolor{cYellow}{HTML}{F9A825}
\definecolor{cBlue}{HTML}{1565C0}
\definecolor{cGray}{HTML}{616161}
\definecolor{lightgreen}{HTML}{E8F5E9}
\definecolor{lightred}{HTML}{FFEBEE}
\definecolor{lightyellow}{HTML}{FFF8E1}
\definecolor{lightblue}{HTML}{E3F2FD}

\newcommand{\pn}{P(N)}
\newcommand{\pa}{P(A)}
\newcommand{\ps}{P(S)}
\newcommand{\piN}{\pi^N}
\newcommand{\piA}{\pi^A}
\newcommand{\piS}{\pi^S}
\newcommand{\Tback}{T_{\text{back}}}
\newcommand{\Talert}{T_{\text{alert}}}
\newcommand{\Tstay}{T_{\text{stay}}}
\newcommand{\Wbase}{W_{\text{base}}}

\newcommand{\recover}{\textcolor{cGreen}{\textbf{RECOVER}}}
\newcommand{\alert}{\textcolor{cRed}{\textbf{ALERT}}}
\newcommand{\inconclusive}{\textcolor{cYellow!80!black}{\textbf{INCONCLUSIVE}}}

\usepackage[breaklinks=true,colorlinks,linkcolor=cBlue,citecolor=cGreen,urlcolor=cBlue,bookmarks=false]{hyperref}

\begin{document}

\title{VIGIL: Verifying Identity via Gated Intermittent Likelihoods for Continuous Biometric Authentication}

\author{
Aldridge Fonseca\\
San Jos\'{e} State University\\
\tt\small ORCID: 0009-0002-0712-6493 \\ \tt\small aldridge.fonseca@sjsu.edu
\and
Udayan Atreya\\
San Jos\'{e} State University\\
\tt\small ORCID: 0009-0001-9509-6839 \\ \tt\small udayan.atreya@sjsu.edu
\and
Amith Kamath Belman\\
San Jos\'{e} State University\\
\tt\small ORCID: 0000-0003-1008-3025 \\ \tt\small amith.kamathbelman@sjsu.edu
\and
Frank Sicong Chen\\
Dartmouth College\\
\tt\small ORCID: 0000-0003-3573-4795 \\ \tt\small frank.chen@dartmouth.edu
}

\maketitle

\begin{abstract}

Continuous multi-modal authentication has emerged as a necessity for securing modern environments against persistent threats. Existing temporal fusion techniques fail to identify a persistent attacker from a genuine user with poor signal strength.
In this study, we propose VIGIL (Verifying Identity via Gated Intermittent Likelihoods for Continuous Biometric Authentication), a highly adaptive continuous authentication framework. We introduce configurable cross-modal fusion with per-modality weighting, enabling operators to select their choice of integration strategy. We improve temporal fusion using dual-state State Transition Machines (STM) with unidirectional transition matrices. A three-zone verification decision model that enables multi-round verification when evidence is inconclusive is used in combination with an adaptive shrinking verification window.
Monotonic decay, backflow elimination and analytical evaluation demonstrate that the proposed framework effectively addresses the limitations of existing approaches and reduces the time to detect intrusions while maintaining high usability for legitimate users.

\end{abstract}

\section{Introduction}
\label{sec:introduction}

Point-of-entry authentication mechanisms are inadequate for modern zero-trust environments \cite{nist2020zerotrust}, where systems need to prevent post-login attacks throughout the session. Continuous authentication mitigates this limitation by verifying the user's identity repeatedly throughout the session and leveraging biometric signals such as face recognition, voice, or keystroke patterns \cite{sitova2016hmog}. Multi-modal biometrics is another mechanism that verifies users by studying multiple signals simultaneously and combining them to authenticate \cite{stylios2021review}. Continuous authentication combined with multi-modal biometrics achieves a high degree of security. However, multi-modal continuous authentication systems encounter multiple challenges in real-world deployments. 

Research has explored Markov-chain \cite{norris2021markov} based frameworks to implement multi-modal continuous authentication systems \cite{ContinousAuthentication} to incorporate the temporal dependency of continuous data streams. State transition machines (STMs) are used to adapt to real-world dynamics by using exponential decay functions to ensure that past states diminish over time \cite{Tuominen_Tweedie_1979}.
However, intermittent data fluctuations and the variable strength of the different signals cause an increase in false alarms in standard Markov based systems. Recent studies have successfully demonstrated how leveraging dual-STM architecture can reduce false alarm rates offering more robust and user-friendly solutions \cite{sspra}. Although existing architectures have implemented temporal fusion with intermittent states and verification windows which add some buffer before the alert is triggered \cite{sspra}, these static approaches raise a trade-off; a short window can trigger false alerts due to momentary disruption in signal, and a long window can increase the time to detect an intrusion. Another limitation is that in the specific case of periods with no biometric observations, the probability of the system remaining in a Normal state inflates due to a backflow in the transition matrices. This backflow grants attackers a free recovery window towards a secure state without requiring biometric evidence.

To mitigate these challenges, this paper proposes a highly adaptive multi-modal authentication framework leveraging the dual-state Markov chain approach. By dynamically adapting the verification window and user-selectable strategies for cross-modal fusion, the proposed architecture minimizes accidental lockouts without compromising security. The primary contributions of this paper are as follows:

\begin{enumerate}
    \item \textbf{Flexible Cross-Modal Fusion:} A modular strategy allowing administrators to select between product rule, weighted sum, and weighted geometric mean \cite{kittler1998combining} providing configurable handling of multi-modal data streams based on specific use cases. 
    \item \textbf{Dual-state Robust Markov Modeling:} A dual-state architecture where STM-1 deals with initial suspicion detection and STM-2 is responsible for alert escalation. More importantly, STM-2 has a unidirectional transition matrix for verification, ensuring that the system remains sensitive to persistent and fluctuating unauthorized attempts, disabling attackers from breaching the system by leveraging the backflow nature of general Markov chains.
    \item \textbf{Adaptive Shrinking Window:} A dynamic verification window during temporal fusion actively shrinking based on time and penalty decay. This allows the system to enforce stricter security boundaries during suspicious states while still keeping the system robust during data fluctuations.
    
\end{enumerate}

The paper is organized as follows: Section II discusses the background and related work, while Section III describes the VIGIL framework. Finally, Section IV concludes our findings and discusses about future results.

\section{Background and Related Work}
\label{sec:related_work}

Continuous authentication systems have shifted towards multi-modal approaches to integrate diverse biometric signals, improving robustness to deal with limited accuracy in high-security environments \cite{ContinousAuthentication}. Despite these advancements, real-world deployment of multi-modal biometric systems has significant challenges; variation in signal quality \cite{abuhamad_survey, baig_survey} and disconnection of modalities \cite{sspra, Ma2021SMILML, wu2026deep} are some of the primary issues when dealing with behavioral biometrics. Furthermore, many traditional frameworks evaluate continuous data streams as isolated events that neglect human behavior and reduce user experience by flagging transient noise \cite{ContinuousAuthenticationforVoiceAssistants}.

Combining continuous data streams from multiple sources requires robust fusion techniques. Multimodal fusion has been widely studied in continuous authentication at different fusion levels. Stylios et al. \cite{STYLIOS2023103363} proposed a feature-level fusion approach that combines touch-gesture and keystroke-dynamics features for mobile continuous authentication, showing that jointly modeling the two modalities can improve security and usability. Ray-Dowling et al. \cite{RAYDOWLING2022102868} evaluated score-level fusion for mobile behavioral biometrics by combining acceleration, gyroscope, and swipe-based authentication scores across public mobile datasets. Fridman et al. \cite{FRIDMAN2015142} proposed a decision-level fusion architecture that integrates multiple behavioral biometric sensors, including keystroke dynamics, mouse movement, and stylometry, for continuous authentication. More recently, Chen et al. \cite{sspra} proposed SSPRA, a two-level fusion framework that combines likelihoods from available modalities at each inspection through vertical-level fusion and incorporates temporal state continuity through horizontal-level fusion. Attrapadung et al. \cite{Attrapadung_fusion} also introduce a two-dimensional fusion technique however they focus more on dynamically selecting a subset of classifiers based on the user's environmental context to optimize the balance between authentication accuracy and device resource consumption.

Feature-level fusion is computationally heavy and struggles with the varied sampling rate of multi-modal sensors \cite{patel2016continuous, ross2003information}. Consequently, score-level fusion acts as a standard method using two dimensions: vertical level fusion and horizontal fusion \cite{sspra}. Chen et al. \cite{sspra} show that vertical fusion allows integration of biometric scores across multiple modalities, while horizontal fusion deals with the temporal integration evaluating these fused scores for a time window to determine the final state of the system (compromised or not compromised).

Although foundational state-space frameworks perform robustly with standard modalities \cite{alotaibi2021deep, hintze2021continuous}, fluctuating signals and volatile environments where prolonged sensor dropouts are common reveal limitations of these frameworks. Vertical fusion in these frameworks is implemented using a static window duration ($W_{base}$) creating a vulnerability for persistent attackers. An impostor can trigger suspense, wait out the verification window while the system resets, and repeat again. Such sophisticated probing grants an attacker the full verification duration every time, delaying the Time to Correct Alarm (TCA), as the system does not account for repeated suspicion. Furthermore, in continuous Markov models, transition probabilities are designed to decay over time reflecting the depletion of past state influence. However, mathematical analysis indicates that periods of complete vacancy of observations lead to structural backflows from Suspense to Normal states, granting recovery over the passage of time without actual biometric evidence.

State space frameworks are also restricted by rigid vertical fusion design relying on fusion techniques such as product rule:
\begin{equation}
    L = \prod_{i=1}^{M} P(s_i|H)
\end{equation}
where $H$ represents the hypothesis (Genuine or Imposter). The product rule is mathematically unforgiving; if a single modality performs poorly (near zero probability due to noise or disconnection), the entire fused likelihood collapses to zero. This ``weakest link'' behavior causes false alarms in real-world scenarios where erratic behavioral sensors may experience partial outages or temporary noise bursts, even if other sensors are reporting strong positive matches.

\section{The VIGIL Framework}
\label{sec:vigil_framework}

\begin{table}[ht]
\caption{Notation used in the VIGIL framework.}
\label{tab:notation}
\centering
\small
\begin{tabular}{@{}l p{5.8cm}@{}}
\toprule
\textbf{Symbol} & \textbf{Definition} \\
\midrule
\multicolumn{2}{@{}l}{\textit{System modes \& states}} \\
$P_1$, $P_2$ & Cruise mode (STM1) and verification mode (STM2) \\
$N$, $S$, $A$ & Normal, Suspense, Alert states \\
$\Omega_1$, $\Omega_2$ & State spaces: $\{N,S\}$ for $P_1$; $\{N,S,A\}$ for $P_2$ \\[3pt]
\multicolumn{2}{@{}l}{\textit{Cross-modal fusion}} \\
$M_j$ & Biometric modality $j$ \\
$\mathcal{A}_t$ & Set of active modalities at time $t$ \\
$P_j^N$, $P_j^{\lnot N}$ & Per-modality intra-/inter-class likelihoods \\
$L_N$, $L_{\lnot N}$ & Fused likelihoods (normal / not-normal) \\
$\hat{w}_j$ & Normalized weight for modality $j$ \\[3pt]
\multicolumn{2}{@{}l}{\textit{Temporal fusion}} \\
$P^N$, $P^S$, $P^A$ & State probabilities (posterior) \\
$\pi^N$, $\pi^S$, $\pi^A$ & State probabilities (prior, after temporal fusion) \\
$p(\Delta t)$ & STM1 decay function, half-life $\tau_p$ \\
$q(\Delta t)$ & STM2 decay function, half-life $\tau_q$ \\[3pt]
\multicolumn{2}{@{}l}{\textit{Decision thresholds}} \\
$T_{stay}$ & P1 stay threshold (below $\to$ enter P2) \\
$T_{back}$ & P2 recovery threshold ($P^N_{post} \geq T_{back}$) \\
$T_{alert}$ & P2 alert threshold ($P^N_{post} \leq T_{alert}$) \\[3pt]
\multicolumn{2}{@{}l}{\textit{Adaptive window}} \\
$W_{base}$ & Base verification window duration \\
$W_{\min}$ & Minimum window duration \\
$\delta$ & Penalty increment per P2 entry \\
$\lambda$ & Penalty decay rate in Normal \\
\bottomrule
\end{tabular}
\end{table}

\textbf{VIGIL} (\textbf{V}erifying \textbf{I}dentity via \textbf{G}ated \textbf{I}ntermittent \textbf{L}ikelihoods) is a dual-state Markov framework for continuous authentication on intermittent biometric signals. Table~\ref{tab:notation} summarizes the notation used throughout. Building on the two-level fusion architecture and dual-STM mechanism of SSPRA~\cite{sspra}, VIGIL introduces four design principles:

\begin{enumerate}
    \item \textbf{Layered Cross-Modal Fusion.} A configurable, per-subset fusion strategy with per-modality weighting, which allows the system to adapt to the statistical properties of the available modalities (Section~\ref{sec:vertical_fusion}).
    \item \textbf{Unidirectional Temporal Fusion.} Transition matrices in both STM1 and STM2 ensure that only cross-modal fusion (processing of actual biometric evidence) can increase $P(N)$. In the absence of observations, $P(N)$ will decay monotonically (Section~\ref{sec:horizontal_fusion}).
    \item \textbf{Three-Zone Verification.} A decision model in STM2 that is used to distinguish between recovery, alert, and inconclusive outcomes, which enables multi-round verification when evidence is ambiguous (Section~\ref{sec:three_zone}).
    \item \textbf{Adaptive Window Shrinking.} A penalty-based mechanism that reduces the verification window for repeated suspense entries. This adds increasing pressure on threats which are persistent while simultaneously allowing legitimate users to recover naturally (Section~\ref{sec:adaptive_window}).
\end{enumerate}

\begin{figure*}[t]
\centering
\begin{tikzpicture}[
    font=\small,
    >=Stealth,
    block/.style={draw=#1!55!black, fill=#1!10, rounded corners=4pt, thick,
        align=center},
    sb/.style={draw=blue!55!black, fill=blue!12, rounded corners=2pt, thick,
        minimum width=0.7cm, minimum height=0.42cm, font=\scriptsize},
    ob/.style={draw=#1!60!black, fill=#1!25, rounded corners=6pt, thick,
        minimum width=1.3cm, minimum height=0.6cm, font=\small\bfseries,
        align=center},
    mc/.style={draw=black!70, circle, minimum size=0.34cm, font=\tiny\bfseries,
        inner sep=0pt, thick},
    farr/.style={->, very thick, >=Stealth},
    miniarr/.style={->, thick, >=Stealth, black!65},
    al/.style={font=\scriptsize, fill=white, inner sep=1.5pt},
    ttl/.style={font=\small\bfseries},
    sub/.style={font=\scriptsize},
    mtxt/.style={font=\tiny},
]

\node[sb] (m1) at (0, 1.0)  {$M_1$};
\node[sb] (m2) at (0, 0.45) {$M_2$};
\node[font=\scriptsize, gray] at (0, 0.0) {$\vdots$};
\node[sb] (mk) at (0,-0.45) {$M_k$};
\node[sub, anchor=north] at (0,-0.85) {$\mathcal{A}_t$};

\node[block=blue, minimum width=3.2cm, minimum height=2.4cm] (vf) at (2.9, 0.25) {};
\node[font=\footnotesize\bfseries, anchor=north] at (vf.north) {Cross-Modal Fusion};
\node[mc, fill=blue!25] (vd1) at (2.1,  0.55) {};
\node[mc, fill=blue!25] (vd2) at (2.1,  0.20) {};
\node[mc, fill=blue!25] (vd3) at (2.1, -0.15) {};
\node[mc, fill=blue!55] (vout) at (3.6, 0.20) {};
\draw[miniarr, blue!60] (vd1) -- (vout);
\draw[miniarr, blue!60] (vd2) -- (vout);
\draw[miniarr, blue!60] (vd3) -- (vout);
\node[sub, anchor=south] at (vf.south) {$\hat{w}_j \to L_N,\, L_{\lnot N}$};

\fill[gray!4, rounded corners=6pt, draw=gray!30, thick]
    (4.9, -3.0) rectangle (12.0, 2.5);
\node[ttl, anchor=north west]
    at (5.0, 2.45) {Temporal Fusion \& Decision};

\node[block=green, minimum width=5.6cm, minimum height=1.4cm] (p1) at (8.1, 1.3) {};
\node[ttl, anchor=north] at (p1.north) {P1 \,---\, Cruise};
\node[mc, fill=green!30] (p1n) at (7.7, 1.1) {N};
\node[mc, fill=yellow!30] (p1s) at (8.5, 1.1) {S};
\draw[miniarr] (p1n) -- (p1s);
\path (p1n) edge[->, thick, black!65, loop above, looseness=8] (p1n);
\path (p1s) edge[->, thick, black!65, loop above, looseness=8] (p1s);

\node[block=orange, minimum width=5.6cm, minimum height=1.7cm] (p2) at (8.1, -1.5) {};
\node[ttl, anchor=north] at (p2.north) {P2 \,---\, Verification};
\node[mc, fill=green!30] (p2n) at (5.55, -1.55) {N};
\node[mc, fill=yellow!30] (p2s) at (6.15, -1.55) {S};
\node[mc, fill=red!30] (p2a) at (6.75, -1.55) {A};
\draw[miniarr] (p2n) -- (p2s);
\draw[miniarr] (p2s) -- (p2a);
\fill[green!45]  (7.45, -1.65) rectangle (7.78, -1.45);
\fill[yellow!50] (7.78, -1.65) rectangle (8.11, -1.45);
\fill[red!45]    (8.11, -1.65) rectangle (8.44, -1.45);
\draw[gray!60, thin] (7.45, -1.65) rectangle (8.44, -1.45);
\node[mtxt, anchor=north] at (7.95, -1.7) {3-zone};
\fill[orange!25] (8.85, -1.65) rectangle (9.55, -1.45);
\fill[orange!40] (9.6, -1.65)  rectangle (10.1, -1.45);
\fill[orange!60] (10.15,-1.65) rectangle (10.45,-1.45);
\draw[gray!60, thin] (8.85, -1.65) rectangle (10.45, -1.45);
\node[mtxt, anchor=north] at (9.65, -1.7) {adaptive window};
\node[sub, anchor=south, orange!60!black] at (p2.south)
    {$\circlearrowleft$ multi-round};

\node[ob=green] (ok)    at (13.1,  1.3) {RECOVER};
\node[ob=red]   (alert) at (13.1, -1.5) {ALERT};

\draw[farr, blue!50] (m1.east) -- (vf.west |- m1);
\draw[farr, blue!50] (m2.east) -- (vf.west |- m2);
\draw[farr, blue!50] (mk.east) -- (vf.west |- mk);

\draw[farr, black!55] (vf.east) -- (4.9, 0.25);

\draw[farr, green!55!black] (p1.east) -- (ok.west);

\draw[farr, red!60!black] (p2.east) -- (alert.west);

\draw[farr, red!65!black, rounded corners=2pt]
    ([xshift=-12pt]p1.south) -- node[left, al] {$P(N) < T_{stay}$}
    ([xshift=-12pt]p2.north);

\draw[farr, green!55!black, rounded corners=2pt]
    ([xshift=12pt]p2.north) -- node[right, al] {$P(N) \geq T_{back}$}
    ([xshift=12pt]p1.south);

\end{tikzpicture}
\caption{VIGIL system architecture. Active modalities $\mathcal{A}_t$ feed Cross-Modal Fusion to produce $L_N, L_{\lnot N}$. Temporal fusion operates in two modes: P1 (cruise, unidirectional decay) and P2 (verification, with three-zone decisions and adaptive window across multiple rounds). Mode transitions are driven by $T_{stay}$ (escalate) and $T_{back}$ (recover); P2 escalates to ALERT on $\leq T_{alert}$ or after maximum rounds.}
\label{fig:vigil_arch}
\end{figure*}
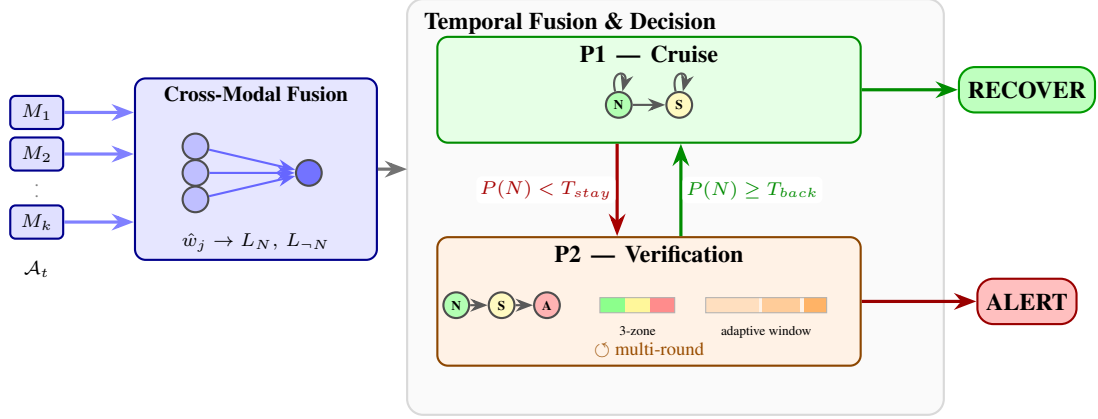

Figure~\ref{fig:vigil_arch} provides an overview of the complete VIGIL pipeline.

\subsection{Layered Cross-Modal Fusion}
\label{sec:vertical_fusion}

In multi-modal fusion, different modalities have varying levels of reliability, and classifiers may exhibit different performance. A standard product rule:
\begin{equation}
P(\mathcal{M}_t \mid s_t) = \prod_{j \in \mathcal{A}_t} P(m_t^j \mid s_t)
\label{eq:original_vertical}
\end{equation}
assigns equal weight to every modality, so a single weak sensor can collapse the fused score. Instead, VIGIL explores alternative fusion strategies with per-modality weighting that account for these differences. Let $P_j^N$ and $P_j^{\lnot N}$ denote the intra-class and inter-class likelihoods for modality~$j$. The primary strategies are given in Table~\ref{tab:fusion_strategies}.

Each $P_j^N$ and $P_j^{\lnot N}$ is obtained by evaluating the modality's score against per-modality class-conditional probability mass functions (PMFs) estimated from labeled validation data, following standard score-level-fusion practice. VIGIL is agnostic to how these likelihoods are produced and it only requires that $P_j^N, P_j^{\lnot N} \in [0,1]$.

\begin{table}[ht]
\caption{VIGIL fusion strategy characteristics. \textsuperscript{*}$\hat{w}_j$ denotes the normalized weight for modality~$j$, satisfying $\sum_{j \in \mathcal{A}_t} \hat{w}_j = 1$.}
\label{tab:fusion_strategies}
\centering
\resizebox{\columnwidth}{!}{
\begin{tabular}{@{}p{2.0cm} p{2.8cm} p{3.4cm}@{}}
\toprule
\textbf{Strategy} & \textbf{Formulation} & \textbf{Behavior} \\
\midrule
Product & $L = \prod_{j} P_j$ & Sensitive to weakest modality; collapses if any $P_j \approx 0$ \\
Weighted Sum\textsuperscript{*} & $L = \sum_{j} \hat{w}_j P_j$ & Tolerates dropouts; linear combination \\
Weighted Geometric Mean\textsuperscript{*} & $L = \prod_{j} P_j^{\hat{w}_j}$ & Multiplicative with damping; balances sensitivity (\textbf{default}) \\
Max Rule & $L = \max_{j} P_j$ & Driven by strongest signal; ignores weak modalities \\
Min Rule & $L = \min_{j} P_j$ & Driven by weakest signal; conservative bound \\
\bottomrule
\end{tabular}
}
\end{table}

VIGIL uses the weighted geometric mean as its default fusion strategy. The geometric mean is chosen as it preserves the multiplicative structure of the product rule. Thus it is able to maintain sensitivity to each modality, while simultaneously preventing a single degraded sensor from making the fused likelihood collapse to zero. When all sensors are equally reliable, setting $\hat{w}_j = 1/|\mathcal{A}_t|$ recovers the unweighted geometric mean. In practice, however, weights will be assigned based on each modality's expected signal quality. This could be determined through validation-set EER or FAR/FRR characteristics.

The strategies above are representative, and VIGIL will work for any function $f: \mathbb{R}^{|\mathcal{A}_t|} \rightarrow \mathbb{R}$ that maps per-modality likelihoods to a non-negative scalar. So users of the system could add domain-specific rules (e.g., trimmed means, Dempster--Shafer) without any modifications to the framework.

\subsubsection{Subset-Configurable Fusion}

VIGIL allows each subset of active modalities $\mathcal{A}_t$ to have its own fusion function and weight vector. Different sensor combinations may have different noise profiles, and thus may require different fusion strategies. At each observation, VIGIL retrieves the configuration for the current $\mathcal{A}_t$; if none is defined, it will use the default fusion function and weight vector. Table~\ref{tab:subset_config} provides an example for a multi-modal system.

Consider a system with the following validation-set Equal Error Rates: $M_1$ (face) EER~=~2\%, $M_2$ (iris) EER~=~1.5\%, $M_3$ (voice) EER~=~8\%, $M_4$ (keystroke) EER~=~12\%, $M_5$ (gait) EER~=~15\%. Weights are determined from inverse EER and are then normalized per active subset. The fusion strategy per subset is chosen based on the noise characteristics of the sensors present. Weighted geometric mean is used when all sensors are reliable. Weighted sum is used when the subset includes noisy or dropout-prone behavioral modalities.

\begin{table}[ht]
\caption{Example subset-configurable fusion for a five-modality system. Weights are derived from inverse EER normalized per active subset.}
\label{tab:subset_config}
\centering
\resizebox{\columnwidth}{!}{%
\begin{tabular}{@{}l l l p{3.2cm}@{}}
\toprule
\textbf{Active Subset} & \textbf{Strategy} & \textbf{Weights} & \textbf{Rationale} \\
\midrule
$\{M_1{\text{--}}M_5\}$ & Wtd.\ Geom. & $(0.35, 0.46, 0.09, 0.06, 0.05)$ & All available; geometric preserves structure \\
$\{M_1, M_2, M_3\}$ & Wtd.\ Geom. & $(0.38, 0.51, 0.10)$ & Three reliable sensors \\
$\{M_1, M_3, M_4\}$ & Wtd.\ Sum & $(0.71, 0.18, 0.12)$ & Keystroke noisy; sum prevents collapse \\
$\{M_3, M_4, M_5\}$ & Wtd.\ Sum & $(0.46, 0.30, 0.24)$ & All behavioral; tolerates dropouts \\
$\{M_1, M_2\}$ & Wtd.\ Geom. & $(0.43, 0.57)$ & Physiological and reliable \\
$\{M_3, M_5\}$ & Wtd.\ Sum & $(0.65, 0.35)$ & Behavioral and dropout-prone \\
$\{M_i\}$ & Product & $(1.0)$ & No weighting needed \\
\bottomrule
\end{tabular}%
}
\end{table}

\subsection{Unidirectional Temporal Fusion}
\label{sec:horizontal_fusion}

A core principle of VIGIL is that only cross-modal fusion, which is the processing of actual biometric observations, may increase $P(N)$. Without observations, $P(N)$ decays monotonically. VIGIL is able to enforce this design via unidirectional transitions in both STM1 and STM2.

The temporal-fusion decay functions in VIGIL follow the exponential half-life form $f(\Delta t) = e^{-\frac{\ln 2}{\tau} \cdot \Delta t}$, where $\tau$ is a configurable half-life parameter. VIGIL uses two decay functions:
\begin{itemize}
    \item $p(\Delta t)$ with half-life $\tau_p$: this governs Normal-to-Suspense decay during cruise mode (STM1).
    \item $q(\Delta t)$ with half-life $\tau_q$: this governs Normal-to-Alert decay during verification mode (STM2).
\end{itemize}
Separating $p$ and $q$ allows the system to decay at different rates depending on the mode.

\subsubsection{VIGIL STM1 Design}
\label{sec:stm1_design}

During cruise mode (P1), VIGIL uses the following transition matrix for temporal fusion:
\begin{equation}
T_{P1}^{\text{VIGIL}} = \begin{pmatrix} p & 1-p & 0 \\ 0 & 1 & 0 \\ 0 & 0 & 1 \end{pmatrix}
\label{eq:stm1_vigil}
\end{equation}
where the decay function $p(\Delta t) = e^{-\frac{\ln 2}{\tau_p} \cdot \Delta t}$ will govern the rate at which $P(N)$ decays with half-life $\tau_p$. The prior update under this model is given as:
\begin{align}
\pi^N_t &= p \cdot P^N_{t-1} \label{eq:p1_vigil_n} \\
\pi^S_t &= (1-p) \cdot P^N_{t-1} + P^S_{t-1} \label{eq:p1_vigil_s} \\
\pi^A_t &= P^A_{t-1} \nonumber
\end{align}

Suspense is absorbing during temporal fusion; this means that any probability mass leaving Normal cannot return without evidence. This gives VIGIL a strict monotonic decay under no observation conditions:
\begin{equation}
\pi^N_t = p(\Delta t) \cdot P^N_{t-1} < P^N_{t-1} \quad \text{for all } \Delta t > 0
\label{eq:p1_monotonic}
\end{equation}

\subsubsection{Comparison with Baseline SSPRA}
\label{sec:stm1_comparison}

For reference, the Baseline SSPRA STM1 transition matrix includes an $S \to N$ path:
\begin{equation}
T_{P1}^{\text{base}} = \begin{pmatrix} p & 1-p & 0 \\ u & 1-u & 0 \\ 0 & 0 & 1 \end{pmatrix}
\label{eq:stm1_baseline}
\end{equation}
where $u(\Delta t) = e^{-\frac{\ln 2}{\tau_u} \cdot \Delta t}$ models recovery from Suspense to Normal during temporal fusion. The corresponding prior update is:
\begin{align}
\pi^N_t &= p \cdot P^N_{t-1} + u \cdot P^S_{t-1} \label{eq:p1_base_n} \\
\pi^S_t &= (1-p) \cdot P^N_{t-1} + (1-u) \cdot P^S_{t-1} \label{eq:p1_base_s}
\end{align}

\begin{figure}[ht]
\centering
\begin{tikzpicture}[
    >=Stealth,
    state/.style={draw, circle, minimum size=1.1cm, font=\scriptsize\bfseries, thick},
    every edge/.style={draw, thick, ->, >=Stealth},
    lbl/.style={font=\scriptsize, fill=white, inner sep=1pt},
]
    \node[state, fill=gray!8, minimum size=0.9cm, font=\tiny\bfseries] (N1) at (0, 0) {N};
    \node[state, fill=gray!8, minimum size=0.9cm, font=\tiny\bfseries] (S1) at (2.0, 0) {S};

    \path (N1) edge[loop above, looseness=8] node[lbl, above=2pt] {$p$} (N1);
    \path (N1) edge[bend left=15] node[lbl, above] {$1\!-\!p$} (S1);
    \path (S1) edge[thick] node[lbl, below] {$u$} (N1);
    \path (S1) edge[loop above, looseness=8] node[lbl, above=2pt] {$1\!-\!u$} (S1);

    \node[font=\scriptsize\bfseries, anchor=north] at (1.0, -0.9) {(a) Baseline SSPRA};

    \node[state, fill=green!20] (N2) at (4.0, 0) {N};
    \node[state, fill=yellow!15] (S2) at (6.0, 0) {S};

    \path (N2) edge[loop above, looseness=8] node[lbl, above=2pt] {$p$} (N2);
    \path (N2) edge[bend left=15] node[lbl, above] {$1\!-\!p$} (S2);
    \path (S2) edge[loop above, looseness=8] node[lbl, above=2pt] {$1$} (S2);

    \node[font=\scriptsize\bfseries, anchor=north] at (5.0, -0.9) {(b) VIGIL};
\end{tikzpicture}
\caption{P1 state transition diagrams. (a)~Baseline SSPRA includes an $S \to N$ path governed by $u$, enabling backflow without observations. (b)~VIGIL makes Suspense absorbing; $P(N)$ can only decrease during temporal fusion.}
\label{fig:stm1_comparison}
\end{figure}
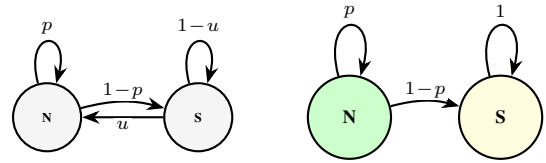

The $u \cdot P^S_{t-1}$ term transfers probability from Suspense back to Normal without biometric evidence. The net change in $P(N)$ is:
\begin{equation}
\Delta P(N) = \underbrace{u \cdot P^S_{t-1}}_{\text{backflow into } N} - \underbrace{(1-p) \cdot P^N_{t-1}}_{\text{decay out of } N}
\label{eq:net_change}
\end{equation}
Whenever $\Delta P(N) > 0$, $P(N)$ increases above its previous value in spite of time passing and there being no new observations; even otherwise, the backflow will still elevate $P(N)$ relative to the unidirectional model (Figure~\ref{fig:stm1_comparison}).

\medskip
\noindent\textbf{Backflow threshold.} Substituting $P^S = 1 - P^N$ and solving $P_{\text{new}}(N) > P_{\text{old}}(N)$ gives:
\begin{equation}
P(N) < \frac{u}{1 - p + u}
\label{eq:inflation_threshold}
\end{equation}

The threshold represents the value of $P(N)$ below which the backflow causes an increase rather than a decrease. Table~\ref{tab:threshold} evaluates this threshold for representative intervals.

\begin{table}[ht]
\caption{Backflow threshold for Baseline SSPRA with $\tau_p = 10\text{s}$, $\tau_u = 4\text{s}$. $P(N)$ increases whenever it is below this value.}
\label{tab:threshold}
\centering
\small
\begin{tabular}{@{}c cc c@{}}
\toprule
$\boldsymbol{\Delta t}$ & $\boldsymbol{p}$ & $\boldsymbol{u}$ & Threshold \\
\midrule
0.25s & 0.9828 & 0.9576 & 0.982 \\
0.50s & 0.9659 & 0.9170 & 0.964 \\
1.0s  & 0.9330 & 0.8409 & 0.926 \\
2.0s  & 0.8706 & 0.7071 & 0.845 \\
5.0s  & 0.7071 & 0.4204 & 0.589 \\
10.0s & 0.5000 & 0.1768 & 0.261 \\
\bottomrule
\end{tabular}
\end{table}

\begin{figure}[ht]
\centering
\includegraphics[width=\columnwidth]{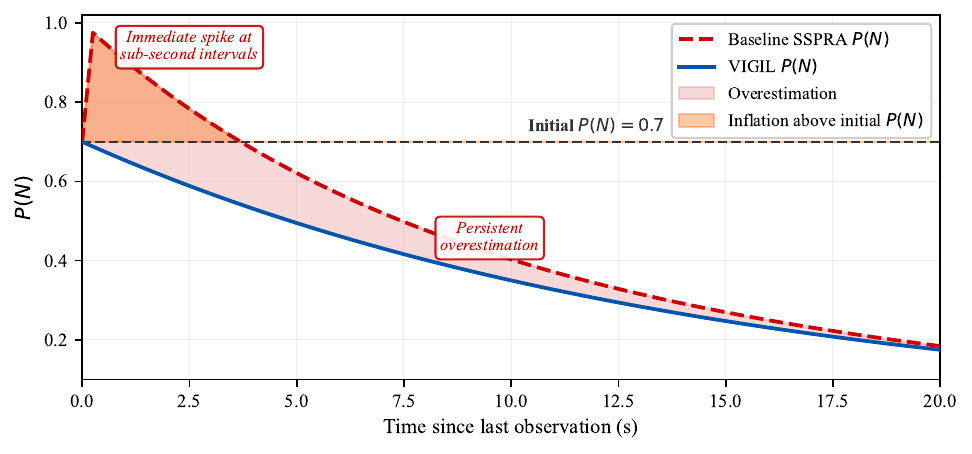}
\caption{STM1 $P(N)$ trajectory with no observations. Baseline SSPRA exhibits a transient spike above the initial $P(N)$ due to $S \to N$ backflow; VIGIL decays monotonically from the first time step.}
\label{fig:stm1_decay}
\end{figure}

At sub-second intervals the threshold approaches 1.0, meaning virtually any realistic $P(N)$ is elevated. Table~\ref{tab:stm1_decay_comparison} and Figure~\ref{fig:stm1_decay} compare trajectories: SSPRA exhibits a transient spike above initial $P(N)$ while VIGIL decays monotonically.

\begin{table}[ht]
\caption{STM1 $P(N)$ with no observations. $\tau_p = 10\text{s}$, $\tau_u = 4\text{s}$. Initial: $P(N) = 0.70$, $P(S) = 0.30$. Bold values exceed the initial $P(N)$.}
\label{tab:stm1_decay_comparison}
\centering
\small
\begin{tabular}{@{}c cc c@{}}
\toprule
$\boldsymbol{\Delta t}$ & \textbf{Baseline SSPRA} & \textbf{VIGIL} & \textbf{Overest.} \\
\midrule
0s (last obs) & 0.7000 & 0.7000 & --- \\
0.25s & \textbf{0.9753} & 0.6880 & +41.7\% \\
0.50s & \textbf{0.9513} & 0.6762 & +40.7\% \\
1.0s  & \textbf{0.9054} & 0.6531 & +38.6\% \\
2.0s  & \textbf{0.8215} & 0.6094 & +34.8\% \\
3.75s & 0.6964 & 0.5398 & +29.0\% \\
5.0s  & 0.6211 & 0.4950 & +25.5\% \\
10.0s & 0.4030 & 0.3500 & +15.1\% \\
20.0s & 0.1844 & 0.1750 & +5.4\% \\
\bottomrule
\end{tabular}
\end{table}

\subsection{Adaptive Multi-Round Verification}
\label{sec:verification_mode}

When $P(N)$ drops below the stay threshold $T_{stay}$ during cruise mode, the system will transition to verification mode (P2). VIGIL's P2 extends the dual-STM mechanism with key features, which are its unidirectional STM2 transitions, a three-zone decision model, and adaptive window shrinking.

\subsubsection{VIGIL STM2 Design}
\label{sec:stm2_design}

At P2 entry, any Suspense state probability mass from P1 is directly transferred to Alert. This gives a starting state $[P^N, 0, P^A]$. VIGIL's STM2 uses the separate decay function $q(\Delta t) = e^{-\frac{\ln 2}{\tau_q} \cdot \Delta t}$ with half-life $\tau_q$. This allows the verification-mode decay rate to differ from cruise mode. The transition matrix during temporal fusion is:
\begin{equation}
T_{P2}^{\text{VIGIL}} = \begin{pmatrix} q & 0 & 1-q \\ 0 & 0 & 1 \\ 0 & 0 & 1 \end{pmatrix}
\label{eq:stm2_vigil}
\end{equation}

The prior update becomes:
\begin{align}
\pi^N_t &= q \cdot P^N_{t-1} \label{eq:p2_vigil_n} \\
\pi^A_t &= (1-q) \cdot P^N_{t-1} + P^A_{t-1} \label{eq:p2_vigil_a} \\
\pi^S_t &= 0 \nonumber
\end{align}

Similar to STM1, $P(N)$ decays monotonically without observations; only cross-modal fusion can restore it.

\subsubsection{Comparison with Baseline SSPRA}
\label{sec:stm2_comparison}

For reference, the Baseline SSPRA STM2 transition matrix includes an $S \to N$ path:
\begin{equation}
T_{P2}^{\text{base}} = \begin{pmatrix} p & 0 & 1-p \\ h & 0 & 1-h \\ 0 & 0 & 1 \end{pmatrix}
\label{eq:stm2_baseline}
\end{equation}
where $h(\Delta t)$ governs the $S \to N$ backflow in STM2, with prior update:
\begin{align}
\pi^N_t &= p \cdot P^N_{t-1} + h \cdot P^S_{t-1} \label{eq:p2_base_n} \\
\pi^A_t &= (1-p) \cdot P^N_{t-1} + (1-h) \cdot P^S_{t-1} + P^A_{t-1} \label{eq:p2_base_a}
\end{align}

\begin{figure}[ht]
\centering
\begin{tikzpicture}[
    >=Stealth,
    state/.style={draw, circle, minimum size=1.1cm, font=\scriptsize\bfseries, thick},
    every edge/.style={draw, thick, ->, >=Stealth},
    lbl/.style={font=\scriptsize, fill=white, inner sep=1pt},
]
    \node[state, fill=gray!8, minimum size=0.9cm, font=\tiny\bfseries] (N1) at (0, 0) {N};
    \node[state, fill=gray!8, minimum size=0.9cm, font=\tiny\bfseries] (S1) at (2.0, 0) {S};
    \node[state, fill=gray!8, minimum size=0.9cm, font=\tiny\bfseries] (A1) at (1.0, -1.6) {A};

    \path (N1) edge[loop above, looseness=8] node[lbl, above=2pt] {$p$} (N1);
    \path (N1) edge[] node[lbl, left] {$1\!-\!p$} (A1);
    \path (S1) edge[thick] node[lbl, above] {$h$} (N1);
    \path (S1) edge[] node[lbl, right] {$1\!-\!h$} (A1);
    \path (A1) edge[loop below, looseness=8] node[lbl, below=2pt] {$1$} (A1);

    \node[font=\scriptsize\bfseries, anchor=north] at (1.0, -3.0) {(a) Baseline SSPRA};

    \node[state, fill=green!20] (N2) at (4.0, 0) {N};
    \node[state, fill=yellow!15] (S2) at (5.8, 0) {S};
    \node[state, fill=red!20] (A2) at (4.9, -1.8) {A};

    \path (N2) edge[loop above, looseness=8] node[lbl, above=2pt] {$q$} (N2);
    \path (N2) edge[] node[lbl, left] {$1\!-\!q$} (A2);
    \path (S2) edge[] node[lbl, right] {$1$} (A2);
    \path (A2) edge[loop right, looseness=8] node[lbl, right=2pt] {$1$} (A2);

    \node[font=\scriptsize\bfseries, anchor=north] at (4.9, -2.6) {(b) VIGIL};
\end{tikzpicture}
\caption{P2 state transition diagrams. (a)~Baseline SSPRA includes an $S \to N$ path governed by $h$, enabling backflow without observations. (b)~VIGIL transfers all Suspense mass to Alert at P2 entry ($S \to A$ with probability~1); only cross-modal fusion can increase $P(N)$.}
\label{fig:stm2_comparison}
\end{figure}
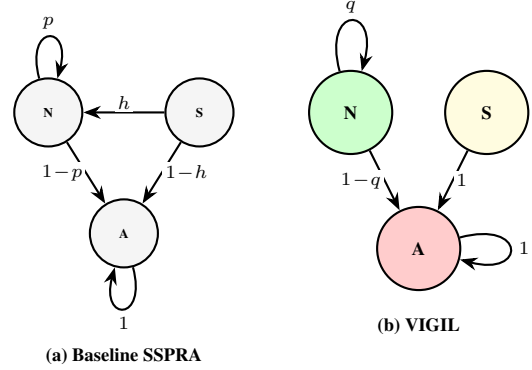

The $h \cdot P^S_{t-1}$ term allows $P(N)$ to rise without biometric evidence, the same backflow effect observed in STM1. With the parameters from Table~\ref{tab:stm1_decay_comparison} and entry state $P(N) = 0.65$, the baseline exceeds $T_{back} = 0.70$ within 0.25\,s, enabling free recovery without any observations. VIGIL's unidirectional STM2 decay prevents this (Figure~\ref{fig:stm2_comparison}).

\subsubsection{Three-Zone Decision Model}
\label{sec:three_zone}

Baseline SSPRA uses a binary P2 decision: recover if $P(N) \geq T_{back}$, otherwise alert. VIGIL replaces this with a three-zone decision model that introduces an inconclusive region, which is the basis of the multi-round verification:

\begin{itemize}
    \item $P^N_{post} \geq T_{back}$: \textbf{Recovery} --- sufficient evidence to confirm the user is genuine. Transfer $P^A \to P^S$ and return to P1.
    \item $P^N_{post} \leq T_{alert}$: \textbf{Alert} --- strong evidence against the user. Lock the session.
    \item $T_{alert} < P^N_{post} < T_{back}$: \textbf{Inconclusive} --- evidence is ambiguous. Shrink window, carry posterior forward, and run another verification round.
    \item No observations received during the window: \textbf{Alert} --- verification failure.
    \item Maximum rounds exceeded while still inconclusive: \textbf{Alert}.
\end{itemize}

\begin{figure}[ht]
\centering
\begin{tikzpicture}[
    scale=0.68, every node/.style={transform shape},
    >=Stealth,
    block/.style={draw, rounded corners, minimum width=2.2cm, minimum height=0.7cm,
                  align=center, font=\scriptsize, thick},
    decision/.style={draw, diamond, aspect=2.5, minimum width=1.8cm,
                     align=center, font=\scriptsize, thick},
    terminal/.style={draw, rounded corners=8pt, minimum width=2.2cm, minimum height=0.7cm,
                     align=center, font=\scriptsize\bfseries, thick},
    arrow/.style={->, thick, >=Stealth},
    every edge quotes/.style={font=\scriptsize, inner sep=2pt},
]
    \node[decision, fill=blue!5] (obs) {Obs.\\received?};

    \node[block, fill=gray!15, left=2.0cm of obs] (start) {P2 Window\\Opens};
    \node[block, fill=blue!10, below=0.8cm of obs] (fusion) {Cross-Modal +\\Temporal Fusion};
    \node[decision, fill=blue!5, below=0.6cm of fusion] (dec) {Evaluate\\$P^N_{post}$};
    \node[block, fill=yellow!20, below=0.8cm of dec] (rewind) {Shrink Window\\Next Round};
    \node[decision, fill=blue!5, below=0.6cm of rewind] (maxr) {Max rounds\\reached?};

    \node[terminal, fill=green!25, left=1.5cm of dec] (recover) {RECOVER\\to P1};
    \node[terminal, fill=red!25, right=1.5cm of dec] (alert) {ALERT\\Lock Session};

    \draw[arrow] (start) -- (obs);

    \draw[arrow] (obs) -- node[left] {Yes} (fusion);
    \draw[arrow] (fusion) -- (dec);
    \draw[arrow] (dec) -- node[left] {Inconclusive} (rewind);
    \draw[arrow] (rewind) -- (maxr);

    \draw[arrow] (dec) -- node[above] {$\geq T_{back}$} (recover);
    \draw[arrow] (dec) -- node[above] {$\leq T_{alert}$} (alert);

    \draw[arrow] (obs) -- node[above] {No} (obs -| alert.north) -- (alert);
    \draw[arrow] (maxr) -- node[below] {Yes} (maxr -| alert.south) -- (alert);

    \draw[arrow, rounded corners=6pt] (maxr.west) -- ++(-4.5,0) |- (start.west);
    \node[font=\scriptsize, fill=white, inner sep=1pt] at ([xshift=-4.5cm, yshift=1.0cm]maxr.west) {No};
\end{tikzpicture}
\caption{VIGIL multi-round P2 verification flow. Three paths lead to alert: no observations, $P^N_{post} \leq T_{alert}$, or max rounds exhausted. Recovery requires $P^N_{post} \geq T_{back}$. Inconclusive rounds shrink the window and repeat.}
\label{fig:p2_decision_flow}
\end{figure}
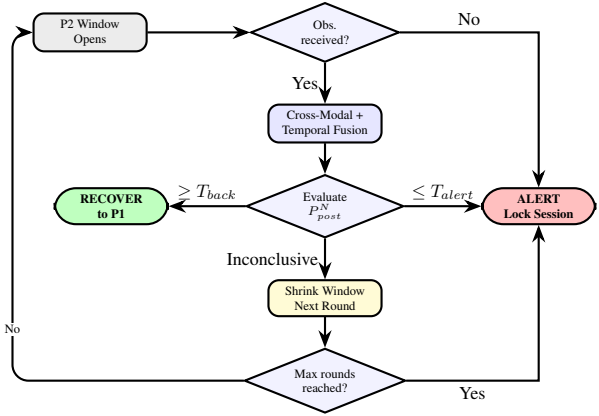

Each additional round compounds pressure: the window shrinks (fewer observations, demanding a stronger signal) and $P(A)$ accumulates (requiring stronger evidence to escape P2 than in the previous round).

\subsubsection{Adaptive Window Shrinking}
\label{sec:adaptive_window}

VIGIL introduces an adaptive penalty mechanism, in which each P2 entry adds a penalty $\delta$ to a running accumulator. The effective window is:
\begin{equation}
W_{\text{current}} = \max\!\bigl(W_{\min},\; W_{base} - \text{penalty}\bigr)
\label{eq:adaptive_window}
\end{equation}
where $W_{\min}$ is the minimum window duration to ensure at least one observation cycle. The penalty decays at rate $\lambda$ while in the Normal state:
\begin{equation}
\text{penalty}(t) = \max\!\bigl(0,\; \text{penalty}_{0} - \lambda \cdot \Delta t_{\text{normal}}\bigr)
\label{eq:penalty_decay}
\end{equation}
where $\Delta t_{\text{normal}}$ is the time in Normal since the last P2 exit. A genuine user who maintains stable behavior sees the penalty disappear. On the other hand, a persistent threat that keeps re-entering P2 in short duration will face a progressively shrinking window.

Table~\ref{tab:adaptive_window} traces a representative session that illustrates both effects.

\begin{table}[ht]
\caption{Adaptive window session timeline ($W_{base}\!=\!5.0$s, $\delta\!=\!1.0$s, $\lambda\!=\!0.1$\,s$^{-1}$, $W_{\min}\!=\!1.0$s). $W_{\text{eff}}$: effective window if P2 is entered at time $t$. \colorbox{red!14}{Red}: P2 entries (penalty increases by $\delta$). \colorbox{green!14}{Green}: Normal periods (penalty decays at $\lambda$).}
\label{tab:adaptive_window}
\centering
\small
\begin{tabular}{@{}r l c c@{}}
\toprule
$\boldsymbol{t}$ & \textbf{Event} & \textbf{Penalty} & $\boldsymbol{W_{\text{eff}}}$ \\
\midrule
\rowcolor{red!8}
$0$\,s & 1st P2 entry & $0.0$ & $5.0$\,s \\
\rowcolor{red!14}
$5$\,s & 2nd P2 entry (immediate) & $1.0$ & $4.0$\,s \\
\rowcolor{red!20}
$9$\,s & 3rd P2 entry (immediate) & $2.0$ & $3.0$\,s \\
\rowcolor{green!14}
$12$\,s & Recovers to Normal & $3.0$ & $2.0$\,s \\
\rowcolor{green!14}
$22$\,s & 10\,s in Normal & $2.0$ & $3.0$\,s \\
\rowcolor{red!14}
$32$\,s & 4th P2 entry & $1.0$ & $4.0$\,s \\
\rowcolor{green!14}
$36$\,s & Recovers to Normal & $2.0$ & $3.0$\,s \\
\rowcolor{green!14}
$56$\,s & 20\,s in Normal (full reset) & $0.0$ & $5.0$\,s \\
\bottomrule
\end{tabular}
\end{table}

At $t\!=\!0$\,s the user enters P2 with a fresh penalty of zero and a full 5.0\,s window. Two immediate re-entries at $t\!=\!5$\,s and $t\!=\!9$\,s increase the penalty to 3.0, shrinking the window to 3.0\,s. The user recovers at $t\!=\!12$\,s and remains in Normal. By $t\!=\!22$\,s (10\,s of Normal), the penalty has decayed from 3.0 to 2.0 and the effective window has partially recovered to 3.0\,s. At $t\!=\!32$\,s (20\,s of Normal), the penalty reaches 1.0 and an isolated 4th P2 entry occurs with a 4.0\,s window. The user recovers again at $t\!=\!36$\,s with penalty 2.0. After another 20\,s of stable Normal behavior ($t\!=\!56$\,s), the penalty decays to zero and the full 5.0\,s window is restored.

\medskip
\noindent\textbf{Parameter rationale.} Table~\ref{tab:p2_params} summarizes the verification parameters and their design rationale.

\begin{table}[ht]
\caption{VIGIL default verification parameters (deployment-configurable).}
\label{tab:p2_params}
\centering
\small
\begin{tabular}{@{}l c p{4.8cm}@{}}
\toprule
\textbf{Parameter} & \textbf{Value} & \textbf{Rationale} \\
\midrule
$W_{base}$ & $5.0$\,s & Default verification window length. Provides sufficient time for multiple biometric observations before a decision is made. \\
$\delta$ & $1.0$\,s & Reduces the window by 1.0\,s per entry. A persistent threat reaches $W_{\min}$ within four consecutive entries, facing meaningful pressure within 3--5 probes. \\
$W_{\min}$ & $1.0$\,s & Lower bound ensures at least one full observation cycle. Prevents zero-length windows. \\
$\lambda$ & $0.1$\,s$^{-1}$ & Penalty decays to zero after ${\approx}\,40$\,s of stable Normal behavior (from max penalty), restoring the full window for legitimate users. \\
$T_{back}$ & $0.70$ & Requires 70\% confidence in genuine identity before recovery. Matches the cruise-mode stay threshold $T_{stay}$. \\
$T_{alert}$ & $0.30$ & Symmetric complement of $T_{back}$: 70\% evidence against the user. Provides a 40 percentage-point inconclusive band. \\
\bottomrule
\end{tabular}
\end{table}

\medskip
\noindent\textbf{Effect of threshold selection on system behavior.}
$T_{back}$ and $T_{alert}$ are deployment-configurable. A higher $T_{back}$ requires stronger evidence for recovery, improving security at the cost of more verification rounds. A higher $T_{alert}$ triggers lockout sooner, reducing attacker dwell time but increasing false alarm risk. The width of the inconclusive band ($T_{back} - T_{alert}$) controls the multi-round behavior; a wider band will get more evidence but delays decisions.
\subsubsection{Multi-Round Behavior}
\label{sec:multi_round}

The three-zone model and adaptive window work together to create compounding pressure across verification rounds. When a round ends as inconclusive, the posterior state carries forward to the next round without reset. This means that if the user cannot produce sufficiently strong biometric evidence, $P(N)$ decreases and $P(A)$ grows with each round. Thus, recovering to cruise mode requires progressively stronger biometric evidence in each subsequent round. This is because cross-modal fusion must now overcome a larger $P(A)$ to push $P(N)$ above $T_{back}$.

If no observations are received during any verification window, the system triggers an immediate alert without waiting for subsequent rounds. This prevents a scenario where an attacker could stall the system by avoiding biometric sensors entirely.

A complete end-to-end numerical walkthrough covering P1 cruise, the P2 transition, and all  verification scenarios is provided in the supplementary material. Figure~\ref{fig:multi_round_timeline} illustrates three representative scenarios.

\begin{figure}[ht]
\centering
\begin{tikzpicture}[x=0.28cm, y=0.7cm,
    rlbl/.style={font=\tiny, anchor=south, inner sep=1pt},
    wlbl/.style={font=\tiny, anchor=north, inner sep=1pt},
    olbl/.style={font=\tiny\bfseries, anchor=west, inner sep=2pt},
    slbl/.style={font=\tiny\bfseries, anchor=east, inner sep=3pt},
    rnd/.style={rounded corners=1pt},
]
\def\h{0.3}
\node[slbl] at (-0.3, 0) {(a)};
\fill[yellow!25,rnd] (0,-\h) rectangle (5,\h);
\draw[rnd] (0,-\h) rectangle (5,\h);
\node[rlbl] at (2.5,\h) {R1};
\node[wlbl] at (2.5,-\h) {5s};
\fill[yellow!25,rnd] (5.35,-\h) rectangle (9.35,\h);
\draw[rnd] (5.35,-\h) rectangle (9.35,\h);
\node[rlbl] at (7.35,\h) {R2};
\node[wlbl] at (7.35,-\h) {4s};
\fill[green!25,rnd] (9.7,-\h) rectangle (12.7,\h);
\draw[rnd] (9.7,-\h) rectangle (12.7,\h);
\node[rlbl] at (11.2,\h) {R3};
\node[wlbl] at (11.2,-\h) {3s};
\node[olbl, green!50!black] at (13.05, 0) {RECOVER};
\def\yb{-1.3}
\node[slbl] at (-0.3, \yb) {(b)};
\fill[yellow!25,rnd] (0,\yb-\h) rectangle (5,\yb+\h);
\draw[rnd] (0,\yb-\h) rectangle (5,\yb+\h);
\node[rlbl] at (2.5,\yb+\h) {R1};
\node[wlbl] at (2.5,\yb-\h) {5s};
\fill[yellow!25,rnd] (5.35,\yb-\h) rectangle (9.35,\yb+\h);
\draw[rnd] (5.35,\yb-\h) rectangle (9.35,\yb+\h);
\node[rlbl] at (7.35,\yb+\h) {R2};
\node[wlbl] at (7.35,\yb-\h) {4s};
\fill[yellow!25,rnd] (9.7,\yb-\h) rectangle (12.7,\yb+\h);
\draw[rnd] (9.7,\yb-\h) rectangle (12.7,\yb+\h);
\node[rlbl] at (11.2,\yb+\h) {R3};
\node[wlbl] at (11.2,\yb-\h) {3s};
\fill[yellow!25,rnd] (13.05,\yb-\h) rectangle (15.05,\yb+\h);
\draw[rnd] (13.05,\yb-\h) rectangle (15.05,\yb+\h);
\node[rlbl] at (14.05,\yb+\h) {R4};
\node[wlbl] at (14.05,\yb-\h) {2s};
\fill[red!25,rnd] (15.4,\yb-\h) rectangle (16.4,\yb+\h);
\draw[rnd] (15.4,\yb-\h) rectangle (16.4,\yb+\h);
\node[rlbl] at (15.9,\yb+\h) {R5};
\node[wlbl] at (15.9,\yb-\h) {1s};
\node[olbl, red!60!black] at (16.75, \yb) {ALERT};
\def\yc{-2.6}
\node[slbl] at (-0.3, \yc) {(c)};
\fill[yellow!25,rnd] (0,\yc-\h) rectangle (5,\yc+\h);
\draw[rnd] (0,\yc-\h) rectangle (5,\yc+\h);
\node[rlbl] at (2.5,\yc+\h) {R1};
\node[wlbl] at (2.5,\yc-\h) {5s};
\fill[green!25,rnd] (5.35,\yc-\h) rectangle (9.35,\yc+\h);
\draw[rnd] (5.35,\yc-\h) rectangle (9.35,\yc+\h);
\node[rlbl] at (7.35,\yc+\h) {R2};
\node[wlbl] at (7.35,\yc-\h) {4s};
\fill[green!8,rnd] (9.7,\yc-\h) rectangle (12.7,\yc+\h);
\draw[dashed, green!40!black, rnd] (9.7,\yc-\h) rectangle (12.7,\yc+\h);
\node[font=\tiny, green!40!black] at (11.2, \yc) {20s Normal};
\fill[green!25,rnd] (13.05,\yc-\h) rectangle (18.05,\yc+\h);
\draw[rnd] (13.05,\yc-\h) rectangle (18.05,\yc+\h);
\node[rlbl] at (15.55,\yc+\h) {R1};
\node[wlbl] at (15.55,\yc-\h) {5s};
\node[olbl, green!50!black] at (18.4, \yc) {RECOVER};
\end{tikzpicture}
\caption{Multi-round P2 scenarios ($W_{base}\!=\!5.0$\,s, $\delta\!=\!1.0$\,s, $W_{\min}\!=\!1.0$\,s). (a)~Legitimate user recovers after three rounds. (b)~Attacker exhausts max rounds; forced alert at $W_{\min}$. (c)~User recovers in round~2, penalty decays during 20\,s in Normal, second P2 entry uses full restored window. \colorbox{yellow!25}{Yellow}: inconclusive. \colorbox{green!25}{Green}: recover. \colorbox{red!25}{Red}: alert. Dashed: Normal period.}
\label{fig:multi_round_timeline}
\end{figure}
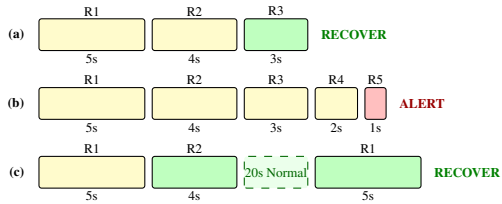

The legitimate user recovers in round~3 because a strong biometric signal overpowers the $P(A)$ accumulation. The attacker, unable to produce strong scores, sees $P(A)$ grow across rounds while the window shrinks from 5.0\,s to 4.0\,s to 3.0\,s to 2.0\,s to 1.0\,s. By round~5, even borderline evidence is insufficient, and the compounding pressure forces a decisive outcome.

\section{Conclusion}
\label{sec:conclusion}

This paper presented VIGIL, a highly adaptive continuous authentication framework built on a dual-state Markov model for intermittent biometric signals. The framework addresses three key challenges. First, flexible cross-modal fusion enabling operators to select the fusion rule with per-modality weighting based on security requirements and use case. Second, a dual-state robust Markov modeling with unidirectional transition matrices in both state machines, which prevent backflow from suspense to normal states. Third, an adaptive shrinking verification window which is paired with a three-zone decision model across multiple verification rounds. All parameters in the framework including fusion rules, thresholds, window duration, and decay rates are configurable. This allows operators to tune the trade-off between detection speed and false alarm rate for their specific environment.

The current evaluation is analytical; VIGIL's properties, such as monotonic decay under no observations, are established through mathematical derivation rather than empirical measurement. Per-modality likelihoods and fusion weights are derived from validation data and are not measured end-to-end within the framework. Additionally, all decay functions use exponential half-life models and alternative decay profiles have not been explored. Empirical validation on multimodal datasets is a natural next step.

Future work could explore extending the framework to support online learning of per-modality weights. This could enable adaptation based on observed sensor reliability in a session. Automated parameter tuning methods could adapt framework parameters to specific deployment environments without any manual configuration. Additionally, empirical evaluation using real multimodal biometric datasets \cite{KHOURY20141147, BANCA} to measure end-to-end FAR/FRR under sensor dropout conditions would validate the framework presented in this work.

{\small
\bibliographystyle{ieee}
\bibliography{references}
}

\clearpage
\onecolumn
\appendix
\setcounter{table}{0}
\renewcommand{\thetable}{S\arabic{table}}
\setcounter{figure}{0}
\renewcommand{\thefigure}{S\arabic{figure}}

\begin{center}
{\LARGE\bfseries Supplementary Material}\\[6pt]
{\large VIGIL: Verifying Identity via Gated Intermittent Likelihoods\\for Continuous Biometric Authentication}\\[12pt]
\end{center}

\noindent This supplement walks through the VIGIL framework end-to-end with full numerical examples. It covers P1 cruise mode, the P2 transition, and six verification scenarios (including a P2$\to$P1$\to$P2 re-entry cycle). Every fusion subset from Table~3 of the main paper is used at least once.

\section{Parameters and Setup}
\label{sec:params}

\subsection{System Parameters}

\begin{table}[H]
\centering
\caption{VIGIL default verification parameters (deployment-configurable).}
\label{tab:p2_params_supp}
\small
\begin{tabular}{@{}l c p{5.5cm}@{}}
\toprule
\textbf{Parameter} & \textbf{Value} & \textbf{Rationale} \\
\midrule
$W_{base}$ & $5.0$\,s & Default verification window length. \\
$\delta$ & $1.0$\,s & Reduces window by 1.0\,s per P2 entry. \\
$W_{\min}$ & $1.0$\,s & Ensures at least one observation cycle. \\
$\lambda$ & $0.1$\,s$^{-1}$ & Penalty decays to zero after ${\approx}\,40$\,s of Normal. \\
$T_{back}$ & $0.70$ & Recovery threshold (matches $T_{stay}$). \\
$T_{alert}$ & $0.30$ & Alert threshold (40\,pp inconclusive band). \\
$\tau_p$ & $10.0$\,s & P1 (STM1) decay half-life. \\
$\tau_q$ & $20.0$\,s & P2 (STM2) decay half-life. \\
Max rounds & $5$ & Maximum P2 rounds before forced alert. \\
\bottomrule
\end{tabular}
\end{table}

\subsection{Modalities and Fusion Subsets}

Five modalities: $M_1$ (face, EER\,=\,2\%), $M_2$ (iris, 1.5\%), $M_3$ (voice, 8\%), $M_4$ (keystroke, 12\%), $M_5$ (gait, 15\%). Table~\ref{tab:subsets} reproduces the subset-configurable fusion from the main paper. \textbf{All seven subsets are exercised} across this walkthrough; each row indicates where it is used.

\begin{table}[H]
\centering
\caption{Subset-configurable fusion (reproduced from Table~3). Weights are inverse-EER normalized per subset.}
\label{tab:subsets}
\small
\begin{tabular}{@{}l l l l@{}}
\toprule
\textbf{Active Subset} & \textbf{Strategy} & \textbf{Weights} & \textbf{Used In} \\
\midrule
$\{M_1\text{--}M_5\}$ & Wtd.\ Geom. & $(0.35, 0.46, 0.09, 0.06, 0.05)$ & Obs\,1 \\
$\{M_1, M_2, M_3\}$ & Wtd.\ Geom. & $(0.38, 0.51, 0.10)$ & Sc.\,A \\
$\{M_1, M_3, M_4\}$ & Wtd.\ Sum & $(0.71, 0.18, 0.12)$ & Sc.\,B R2 \\
$\{M_3, M_4, M_5\}$ & Wtd.\ Sum & $(0.46, 0.30, 0.24)$ & Obs\,3, Sc.\,C R3 \\
$\{M_1, M_2\}$ & Wtd.\ Geom. & $(0.43, 0.57)$ & Obs\,2, Sc.\,F \\
$\{M_3, M_5\}$ & Wtd.\ Sum & $(0.65, 0.35)$ & Sc.\,B R1, Sc.\,C R2 \\
$\{M_i\}$ (single) & Product & $(1.0)$ & Sc.\,C R1 \\
\bottomrule
\end{tabular}
\end{table}

\subsection{Key Equations}

\noindent\textbf{Decay function:}
$f(\Delta t) = e^{-\frac{\ln 2}{\tau} \cdot \Delta t}$

\noindent\textbf{STM1 temporal fusion} (P1, $p = f(\Delta t)$ with $\tau_p$):
\[
\piN = p \cdot \pn_{\text{prev}}, \qquad \piS = (1-p) \cdot \pn_{\text{prev}} + \ps_{\text{prev}}
\]

\noindent\textbf{STM2 temporal fusion} (P2, $q = f(\Delta t)$ with $\tau_q$):
\[
\piN = q \cdot \pn_{\text{prev}}, \qquad \piA = (1-q) \cdot \pn_{\text{prev}} + \pa_{\text{prev}}
\]

\noindent\textbf{Bayesian update:}
$\displaystyle P(N)_{\text{post}} = \frac{L_N \cdot \piN}{L_N \cdot \piN + L_{\lnot N} \cdot \piA}$

\section{Session Timeline}
\label{sec:timeline}

\begin{figure}[H]
\centering
\resizebox{\textwidth}{!}{%
\begin{tikzpicture}[>=Stealth, rnd/.style={rounded corners=2pt}]
\def\xs{0.35}
\def\rh{0.28}
\def\sc{7}
\def\yP{0.3}\def\yA{-1.0}\def\yB{-2.2}\def\yC{-3.4}\def\yD{-4.6}\def\yE{-5.8}\def\yF{-7.0}
\fill[cBlue!6] (-1.1,\yP-0.55) rectangle (38*\xs+0.5,\yP+0.7);
\node[font=\small\bfseries,cBlue,anchor=east] at (-0.7,\yP){P1};
\foreach \lbl/\yr in {A/\yA,B/\yB,C/\yC,D/\yD,E/\yE,F/\yF}
  {\node[font=\scriptsize\bfseries,anchor=east] at (-0.7,\yr){\lbl};}
\foreach \yr in {\yA,\yB,\yC,\yD,\yE,\yF}
  {\draw[cGray!15](-0.5,{\yr-0.55})--(38*\xs+0.3,{\yr-0.55});}
\draw[cBlue!30,very thick](0,\yP)--(5*\xs,\yP);
\fill[lightblue,rnd](-0.25,\yP-\rh) rectangle(0.25,\yP+\rh);
\draw[rnd,dashed,cBlue!60](-0.25,\yP-\rh) rectangle(0.25,\yP+\rh);
\node[font=\tiny] at (0,\yP+0.05){Obs\,1};
\node[font=\tiny,cGreen,anchor=south] at (0,\yP+\rh+0.02){.9943};
\fill[lightblue,rnd](2*\xs-0.25,\yP-\rh) rectangle(2*\xs+0.25,\yP+\rh);
\draw[rnd,dashed,cBlue!60](2*\xs-0.25,\yP-\rh) rectangle(2*\xs+0.25,\yP+\rh);
\node[font=\tiny] at (2*\xs,\yP+0.05){Obs\,2};
\node[font=\tiny,cGreen,anchor=south] at (2*\xs,\yP+\rh+0.02){.9864};
\fill[lightred!50,rnd](5*\xs-0.25,\yP-\rh) rectangle(5*\xs+0.25,\yP+\rh);
\draw[rnd,dashed,cRed!40](5*\xs-0.25,\yP-\rh) rectangle(5*\xs+0.25,\yP+\rh);
\node[font=\tiny] at (5*\xs,\yP+0.05){Obs\,3};
\node[font=\tiny,cRed,anchor=south] at (5*\xs,\yP+\rh+0.02){.6948};
\fill[cRed!60] (6*\xs,\yP) circle(0.06);
\draw[dashed,cRed!40,thick](6*\xs,\yP-0.2)--(6*\xs,\yF-\rh-0.15);
\node[font=\tiny\bfseries,cRed!60,rotate=90,anchor=south] at ({6*\xs-0.2},{(\yC+\yD)/2}){P2 entry};
\foreach \yr in {\yA,\yB,\yC,\yD,\yE,\yF}
  {\draw[->,cGray!50](6*\xs+0.05,\yr)--(\sc*\xs-0.03,\yr);}
\fill[lightgreen,rnd](\sc*\xs,\yA-\rh) rectangle({(\sc+5)*\xs},\yA+\rh);
\draw[rnd,black!25](\sc*\xs,\yA-\rh) rectangle({(\sc+5)*\xs},\yA+\rh);
\node[font=\scriptsize\bfseries] at ({(\sc+2.5)*\xs},\yA+0.06){R1};
\node[font=\tiny,cGray] at ({(\sc+2.5)*\xs},\yA-0.10){5.0\,s};
\node[font=\tiny,cGreen,anchor=south] at ({(\sc+5)*\xs},\yA+\rh+0.02){.9519};
\node[font=\small\bfseries,cGreen,anchor=west] at ({(\sc+5)*\xs+0.2},\yA){RECOVER};
\fill[lightyellow,rnd](\sc*\xs,\yB-\rh) rectangle({(\sc+5)*\xs},\yB+\rh);
\draw[rnd,black!25](\sc*\xs,\yB-\rh) rectangle({(\sc+5)*\xs},\yB+\rh);
\node[font=\scriptsize\bfseries] at ({(\sc+2.5)*\xs},\yB+0.06){R1};
\node[font=\tiny,cGray] at ({(\sc+2.5)*\xs},\yB-0.10){5.0\,s};
\node[font=\tiny,cYellow!80!black,anchor=south] at ({(\sc+5)*\xs},\yB+\rh+0.02){.5901};
\fill[lightgreen,rnd]({(\sc+5)*\xs},\yB-\rh) rectangle({(\sc+9)*\xs},\yB+\rh);
\draw[rnd,black!25]({(\sc+5)*\xs},\yB-\rh) rectangle({(\sc+9)*\xs},\yB+\rh);
\node[font=\scriptsize\bfseries] at ({(\sc+7)*\xs},\yB+0.06){R2};
\node[font=\tiny,cGray] at ({(\sc+7)*\xs},\yB-0.10){4.0\,s};
\node[font=\tiny,cGreen,anchor=south] at ({(\sc+9)*\xs},\yB+\rh+0.02){.8133};
\node[font=\small\bfseries,cGreen,anchor=west] at ({(\sc+9)*\xs+0.2},\yB){RECOVER};
\fill[lightyellow,rnd](\sc*\xs,\yC-\rh) rectangle({(\sc+5)*\xs},\yC+\rh);
\draw[rnd,black!25](\sc*\xs,\yC-\rh) rectangle({(\sc+5)*\xs},\yC+\rh);
\node[font=\scriptsize\bfseries] at ({(\sc+2.5)*\xs},\yC+0.06){R1};
\node[font=\tiny,cGray] at ({(\sc+2.5)*\xs},\yC-0.10){5.0\,s};
\node[font=\tiny,cYellow!80!black,anchor=south] at ({(\sc+5)*\xs},\yC+\rh+0.02){.4837};
\fill[lightyellow,rnd]({(\sc+5)*\xs},\yC-\rh) rectangle({(\sc+9)*\xs},\yC+\rh);
\draw[rnd,black!25]({(\sc+5)*\xs},\yC-\rh) rectangle({(\sc+9)*\xs},\yC+\rh);
\node[font=\scriptsize\bfseries] at ({(\sc+7)*\xs},\yC+0.06){R2};
\node[font=\tiny,cGray] at ({(\sc+7)*\xs},\yC-0.10){4.0\,s};
\node[font=\tiny,cYellow!80!black,anchor=south] at ({(\sc+9)*\xs},\yC+\rh+0.02){.3321};
\fill[lightred,rnd]({(\sc+9)*\xs},\yC-\rh) rectangle({(\sc+12)*\xs},\yC+\rh);
\draw[rnd,black!25]({(\sc+9)*\xs},\yC-\rh) rectangle({(\sc+12)*\xs},\yC+\rh);
\node[font=\scriptsize\bfseries] at ({(\sc+10.5)*\xs},\yC+0.06){R3};
\node[font=\tiny,cGray] at ({(\sc+10.5)*\xs},\yC-0.10){3.0\,s};
\node[font=\tiny,cRed,anchor=south] at ({(\sc+12)*\xs},\yC+\rh+0.02){.2153};
\node[font=\small\bfseries,cRed,anchor=west] at ({(\sc+12)*\xs+0.2},\yC){ALERT};
\fill[lightred,rnd](\sc*\xs,\yD-\rh) rectangle({(\sc+5)*\xs},\yD+\rh);
\draw[rnd,black!25](\sc*\xs,\yD-\rh) rectangle({(\sc+5)*\xs},\yD+\rh);
\node[font=\scriptsize,cRed!70!black] at ({(\sc+2.5)*\xs},\yD){No observations};
\node[font=\small\bfseries,cRed,anchor=west] at ({(\sc+5)*\xs+0.2},\yD){ALERT};
\fill[lightyellow,rnd](\sc*\xs,\yE-\rh) rectangle({(\sc+5)*\xs},\yE+\rh);
\draw[rnd,black!25](\sc*\xs,\yE-\rh) rectangle({(\sc+5)*\xs},\yE+\rh);
\node[font=\scriptsize\bfseries] at ({(\sc+2.5)*\xs},\yE+0.06){R1};
\node[font=\tiny,cGray] at ({(\sc+2.5)*\xs},\yE-0.10){5.0\,s};
\node[font=\tiny,cYellow!80!black,anchor=south] at ({(\sc+5)*\xs},\yE+\rh+0.02){.5843};
\fill[lightyellow,rnd]({(\sc+5)*\xs},\yE-\rh) rectangle({(\sc+9)*\xs},\yE+\rh);
\draw[rnd,black!25]({(\sc+5)*\xs},\yE-\rh) rectangle({(\sc+9)*\xs},\yE+\rh);
\node[font=\scriptsize\bfseries] at ({(\sc+7)*\xs},\yE+0.06){R2};
\node[font=\tiny,cGray] at ({(\sc+7)*\xs},\yE-0.10){4.0\,s};
\node[font=\tiny,cYellow!80!black,anchor=south] at ({(\sc+9)*\xs},\yE+\rh+0.02){.5086};
\fill[lightyellow,rnd]({(\sc+9)*\xs},\yE-\rh) rectangle({(\sc+12)*\xs},\yE+\rh);
\draw[rnd,black!25]({(\sc+9)*\xs},\yE-\rh) rectangle({(\sc+12)*\xs},\yE+\rh);
\node[font=\scriptsize\bfseries] at ({(\sc+10.5)*\xs},\yE+0.06){R3};
\node[font=\tiny,cGray] at ({(\sc+10.5)*\xs},\yE-0.10){3.0\,s};
\node[font=\tiny,cYellow!80!black,anchor=south] at ({(\sc+12)*\xs},\yE+\rh+0.02){.4584};
\fill[lightyellow,rnd]({(\sc+12)*\xs},\yE-\rh) rectangle({(\sc+14)*\xs},\yE+\rh);
\draw[rnd,black!25]({(\sc+12)*\xs},\yE-\rh) rectangle({(\sc+14)*\xs},\yE+\rh);
\node[font=\tiny\bfseries] at ({(\sc+13)*\xs},\yE+0.06){R4};
\node[font=\tiny,cGray] at ({(\sc+13)*\xs},\yE-0.10){2.0\,s};
\node[font=\tiny,cYellow!80!black,anchor=south] at ({(\sc+14)*\xs},\yE+\rh+0.02){.4277};
\fill[lightred,rnd]({(\sc+14)*\xs},\yE-\rh) rectangle({(\sc+15)*\xs},\yE+\rh);
\draw[rnd,black!25]({(\sc+14)*\xs},\yE-\rh) rectangle({(\sc+15)*\xs},\yE+\rh);
\node[font=\tiny\bfseries] at ({(\sc+14.5)*\xs},\yE+0.06){R5};
\node[font=\tiny,cGray] at ({(\sc+14.5)*\xs},\yE-0.10){1.0\,s};
\node[font=\tiny,cRed,anchor=south] at ({(\sc+15)*\xs},\yE+\rh+0.02){.4131};
\node[font=\small\bfseries,cRed,anchor=west] at ({(\sc+15)*\xs+0.2},\yE){ALERT};
\fill[lightgreen,rnd](\sc*\xs,\yF-\rh) rectangle({(\sc+5)*\xs},\yF+\rh);
\draw[rnd,black!25](\sc*\xs,\yF-\rh) rectangle({(\sc+5)*\xs},\yF+\rh);
\node[font=\scriptsize\bfseries] at ({(\sc+2.5)*\xs},\yF+0.06){R1};
\node[font=\tiny,cGray] at ({(\sc+2.5)*\xs},\yF-0.10){5.0\,s};
\node[font=\tiny,cGreen,anchor=south] at ({(\sc+5)*\xs},\yF+\rh+0.02){.9519};
\fill[lightblue!25,rnd]({(\sc+5)*\xs},\yF-\rh) rectangle({(\sc+25)*\xs},\yF+\rh);
\draw[dashed,cBlue!40,rnd]({(\sc+5)*\xs},\yF-\rh) rectangle({(\sc+25)*\xs},\yF+\rh);
\node[font=\tiny,cBlue!60!black] at ({(\sc+15)*\xs},\yF){P1 cruise (20\,s, penalty decays to 0)};
\fill[lightgreen,rnd]({(\sc+25)*\xs},\yF-\rh) rectangle({(\sc+30)*\xs},\yF+\rh);
\draw[rnd,black!25]({(\sc+25)*\xs},\yF-\rh) rectangle({(\sc+30)*\xs},\yF+\rh);
\node[font=\scriptsize\bfseries] at ({(\sc+27.5)*\xs},\yF+0.06){R1};
\node[font=\tiny,cGray] at ({(\sc+27.5)*\xs},\yF-0.10){5.0\,s};
\node[font=\tiny,cGreen,anchor=south] at ({(\sc+30)*\xs},\yF+\rh+0.02){.9178};
\node[font=\small\bfseries,cGreen,anchor=west] at ({(\sc+30)*\xs+0.2},\yF){RECOVER};
\def\yl{-8.0}
\fill[lightblue,rnd](0,\yl-0.12)rectangle(0.5,\yl+0.12);
\draw[rnd,dashed,cBlue!60](0,\yl-0.12)rectangle(0.5,\yl+0.12);
\node[font=\tiny,anchor=west] at (0.6,\yl){P1 Cruise};
\draw[rnd,black!40](1.8,\yl-0.28)rectangle(11.6,\yl+0.28);
\node[font=\tiny\bfseries,anchor=east] at (3.1,\yl){P2 Verification:};
\fill[lightgreen,rnd](3.2,\yl-0.12)rectangle(3.7,\yl+0.12);
\draw[rnd,black!25](3.2,\yl-0.12)rectangle(3.7,\yl+0.12);
\node[font=\tiny,anchor=west] at (3.8,\yl){Recovery};
\fill[lightyellow,rnd](5.8,\yl-0.12)rectangle(6.3,\yl+0.12);
\draw[rnd,black!25](5.8,\yl-0.12)rectangle(6.3,\yl+0.12);
\node[font=\tiny,anchor=west] at (6.4,\yl){Inconclusive};
\fill[lightred,rnd](8.8,\yl-0.12)rectangle(9.3,\yl+0.12);
\draw[rnd,black!25](8.8,\yl-0.12)rectangle(9.3,\yl+0.12);
\node[font=\tiny,anchor=west] at (9.4,\yl){Alert};
\end{tikzpicture}%
}
\caption{Session overview. Row P1 shows cruise-mode observations; rows A--F each show one P2 scenario progressing left to right. $P(N)$ values annotate key transitions. Scenario~F spans the full re-entry cycle with 20\,s of P1 between two P2 episodes.}
\label{fig:timeline}
\end{figure}
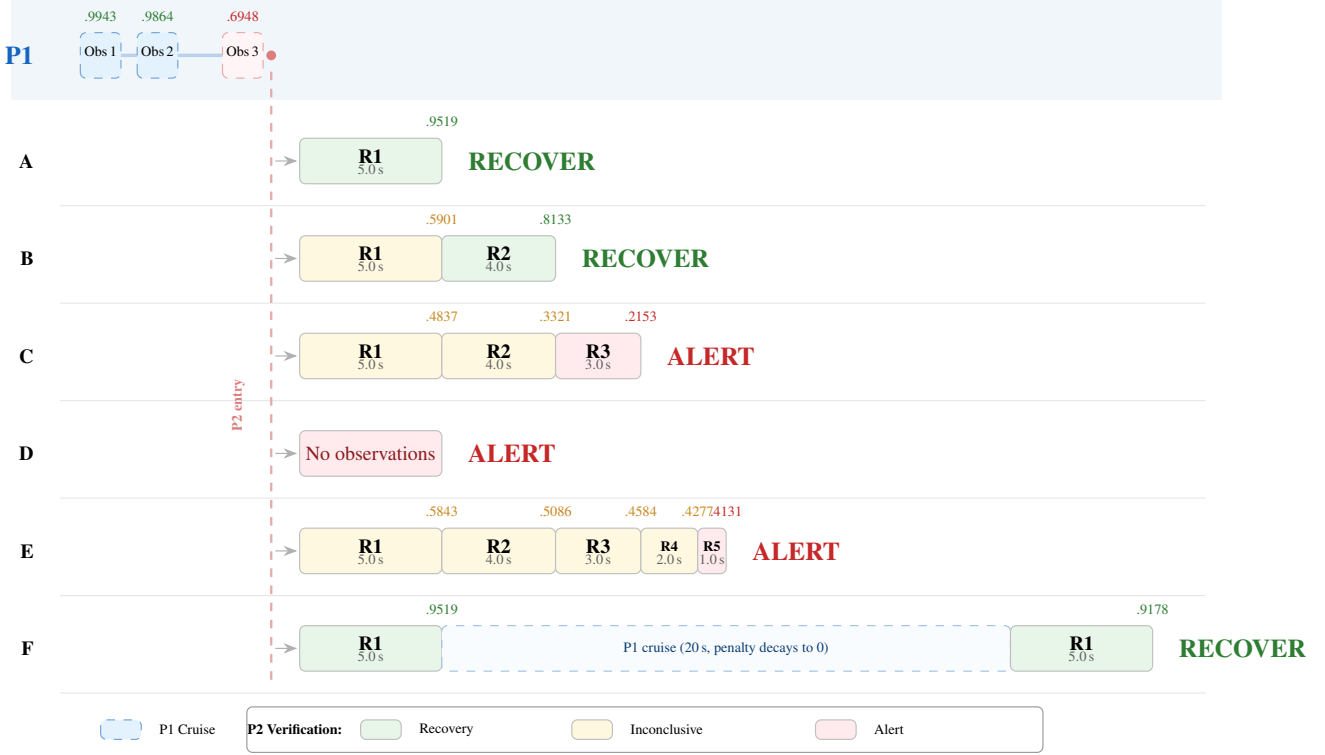

Figure~\ref{fig:timeline} illustrates the complete session. The P1 cruise phase processes three observations at $t=0$, $t=2$\,s, and $t=5$\,s, each using a different fusion subset. After the third observation drops $\pn$ below $\Tstay$, the system enters P2. From this common entry point, six scenarios branch.

\section{Phase 1: P1 Cruise Mode}
\label{sec:p1_cruise}

Session begins with a fresh login. Initial state: $\pn = 0.95$, $\ps = 0.05$.

\subsection{Observation 1 --- \texorpdfstring{$\{M_1\text{--}M_5\}$}{\{M1-M5\}} at \texorpdfstring{$t = 0$}{t=0}\,s}

All five modalities available at login. Weighted Geometric Mean with $\hat{w} = (0.35, 0.46, 0.09, 0.06, 0.05)$.

\begin{enumerate}[itemsep=3pt, parsep=1pt, label=\textbf{\arabic*.}]
\item \textbf{Per-modality scores:}
\begin{center}
\small
\begin{tabular}{@{}l cc@{}}
\toprule
\textbf{Modality} & $P_j^N$ & $P_j^{\lnot N}$ \\
\midrule
$M_1$ (face) & 0.92 & 0.08 \\
$M_2$ (iris) & 0.95 & 0.05 \\
$M_3$ (voice) & 0.55 & 0.45 \\
$M_4$ (keystroke) & 0.50 & 0.50 \\
$M_5$ (gait) & 0.48 & 0.52 \\
\bottomrule
\end{tabular}
\end{center}

\item \textbf{Cross-modal fusion} (Weighted Geometric Mean):
\begin{align*}
L_N &= 0.92^{0.35} \times 0.95^{0.46} \times 0.55^{0.09} \times 0.50^{0.06} \times 0.48^{0.05} = 0.8312 \\
L_{\lnot N} &= 0.08^{0.35} \times 0.05^{0.46} \times 0.45^{0.09} \times 0.50^{0.06} \times 0.52^{0.05} = 0.0900
\end{align*}

\item \textbf{Temporal fusion} ($\Delta t = 0$\,s, so $p = 1.0$):
$\piN = 1.0 \times 0.95 = 0.9500$, \quad $\piS = 0.0 \times 0.95 + 0.05 = 0.0500$

\item \textbf{Bayesian update:}
$P(N)_{\text{post}} = \frac{L_N \cdot \piN}{L_N \cdot \piN + L_{\lnot N} \cdot \piS} = \colorbox{lightgreen}{\,$0.9943$\,}$

\item \textbf{Decision:} $0.9943 \geq \Tstay = 0.70$ $\to$ \textcolor{cGreen}{\textbf{Stay in P1}}.
\end{enumerate}

\subsection{Observation 2 --- \texorpdfstring{$\{M_1, M_2\}$}{\{M1,M2\}} at \texorpdfstring{$t = 2$}{t=2}\,s}

Only physiological sensors available. Weighted Geometric Mean with $\hat{w} = (0.43, 0.57)$.

\begin{enumerate}[itemsep=3pt, parsep=1pt, label=\textbf{\arabic*.}]
\item \textbf{Per-modality scores:}
\begin{center}
\small
\begin{tabular}{@{}l cc@{}}
\toprule
\textbf{Modality} & $P_j^N$ & $P_j^{\lnot N}$ \\
\midrule
$M_1$ (face) & 0.90 & 0.10 \\
$M_2$ (iris) & 0.93 & 0.07 \\
\bottomrule
\end{tabular}
\end{center}

\item \textbf{Cross-modal fusion} (Weighted Geometric Mean):
\begin{align*}
L_N &= 0.90^{0.43} \times 0.93^{0.57} = 0.9170 \\
L_{\lnot N} &= 0.10^{0.43} \times 0.07^{0.57} = 0.0816
\end{align*}

\item \textbf{Temporal fusion} ($\Delta t = 2$\,s, $p = e^{-\frac{\ln 2}{10} \times 2} = 0.8706$):
$\piN = 0.8706 \times 0.9943 = 0.8656$, \quad $\piS = 0.1294 \times 0.9943 + 0.0057 = 0.1344$

\item \textbf{Bayesian update:}
$P(N)_{\text{post}} = \frac{L_N \cdot \piN}{L_N \cdot \piN + L_{\lnot N} \cdot \piS} = \colorbox{lightgreen}{\,$0.9864$\,}$

\item \textbf{Decision:} $0.9864 \geq 0.70$ $\to$ \textcolor{cGreen}{\textbf{Stay in P1}}.
\end{enumerate}

\subsection{Observation 3 --- \texorpdfstring{$\{M_3, M_4, M_5\}$}{\{M3,M4,M5\}} at \texorpdfstring{$t = 5$}{t=5}\,s --- \textcolor{cRed}{Triggers P2}}

Only behavioral sensors active. Noisy environment, poor readings. Weighted Sum with $\hat{w} = (0.46, 0.30, 0.24)$.

\begin{enumerate}[itemsep=3pt, parsep=1pt, label=\textbf{\arabic*.}]
\item \textbf{Per-modality scores:}
\begin{center}
\small
\begin{tabular}{@{}l cc@{}}
\toprule
\textbf{Modality} & $P_j^N$ & $P_j^{\lnot N}$ \\
\midrule
$M_3$ (voice) & 0.40 & 0.60 \\
$M_4$ (keystroke) & 0.35 & 0.65 \\
$M_5$ (gait) & 0.30 & 0.70 \\
\bottomrule
\end{tabular}
\end{center}

\item \textbf{Cross-modal fusion} (Weighted Sum):
\begin{align*}
L_N &= 0.46 \times 0.40 + 0.30 \times 0.35 + 0.24 \times 0.30 = 0.3610 \\
L_{\lnot N} &= 0.46 \times 0.60 + 0.30 \times 0.65 + 0.24 \times 0.70 = 0.6390
\end{align*}

\item \textbf{Temporal fusion} ($\Delta t = 3$\,s since Obs\,2, $p = e^{-\frac{\ln 2}{10} \times 3} = 0.8123$):
$\piN = 0.8123 \times 0.9864 = 0.8012$, \quad $\piS = 0.1877 \times 0.9864 + 0.0136 = 0.1988$

\item \textbf{Bayesian update:}
$P(N)_{\text{post}} = \frac{L_N \cdot \piN}{L_N \cdot \piN + L_{\lnot N} \cdot \piS} = \colorbox{lightred}{\,$0.6948$\,}$

\item \textbf{Decision:} $0.6948 < \Tstay = 0.70$ $\to$ \textcolor{cRed}{\textbf{Enter P2}}.
\end{enumerate}

\section{P2 Entry}
\label{sec:p2_entry}

At P2 entry, Suspense mass transfers to Alert:
\[
\pn = 0.6948, \quad \pa = 1 - 0.6948 = 0.3052
\]

Penalty $= 0$ (first entry). Window $= \Wbase = 5.0$\,s. Six scenarios branch from here.

\section{Scenario A: Immediate Recovery}
\label{sec:scenario_a}

Genuine user provides strong physiological evidence. Subset: $\{M_1, M_2, M_3\}$, Wtd.\ Geom.\ $(0.38, 0.51, 0.10)$.

\subsection{Round 1 \texorpdfstring{$(W = 5.0$}{(W=5.0}\,s)}

\begin{enumerate}[itemsep=3pt, parsep=1pt, label=\textbf{\arabic*.}]
\item \textbf{Entry state:} $\pn = 0.6948$, $\pa = 0.3052$

\item \textbf{Per-modality scores and fusion:}
\begin{center}
\small
\begin{tabular}{@{}l cc@{}}
\toprule
\textbf{Modality} & $P_j^N$ & $P_j^{\lnot N}$ \\
\midrule
$M_1$ (face) & 0.93 & 0.07 \\
$M_2$ (iris) & 0.96 & 0.04 \\
$M_3$ (voice) & 0.60 & 0.40 \\
\bottomrule
\end{tabular}
\end{center}
$L_N = 0.93^{0.38} \times 0.96^{0.51} \times 0.60^{0.10} = 0.9053$, \quad $L_{\lnot N} = 0.07^{0.38} \times 0.04^{0.51} \times 0.40^{0.10} = 0.0643$

\item \textbf{STM2 temporal fusion} ($q = e^{-\frac{\ln 2}{20} \times 5.0} = 0.8409$):
$\piN = 0.8409 \times 0.6948 = 0.5843$, \quad $\piA = 0.1591 \times 0.6948 + 0.3052 = 0.4157$

\item \textbf{Bayesian update:}
$P(N)_{\text{post}} = \frac{L_N \cdot \piN}{L_N \cdot \piN + L_{\lnot N} \cdot \piA} = \colorbox{lightgreen}{\,$0.9519$\,}$

\item \textbf{Decision:} $0.9519 \geq \Tback = 0.70$ $\to$ \recover.
\end{enumerate}

\medskip
\noindent\textit{Outcome:} Return to P1 with $\pn = 0.9519$, $\ps = 0.0481$. Penalty $\to 1.0$\,s.

\section{Scenario B: Multi-Round Recovery}
\label{sec:scenario_b}

Borderline behavioral evidence in R1 (inconclusive), then strong face-dominant evidence in R2.

\subsection{Round 1 \texorpdfstring{$(W = 5.0$}{(W=5.0}\,s) --- \texorpdfstring{$\{M_3, M_5\}$}{\{M3,M5\}}, Wtd.\ Sum}

\begin{enumerate}[itemsep=3pt, parsep=1pt, label=\textbf{\arabic*.}]
\item \textbf{Entry state:} $\pn = 0.6948$, $\pa = 0.3052$

\item \textbf{Per-modality scores and fusion} ($\hat{w} = [0.65, 0.35]$):
\begin{center}
\small
\begin{tabular}{@{}l cc@{}}
\toprule
& $P_j^N$ & $P_j^{\lnot N}$ \\
\midrule
$M_3$ (voice) & 0.52 & 0.48 \\
$M_5$ (gait) & 0.48 & 0.52 \\
\bottomrule
\end{tabular}
\end{center}
$L_N = 0.65 \times 0.52 + 0.35 \times 0.48 = 0.5060$, \quad $L_{\lnot N} = 0.65 \times 0.48 + 0.35 \times 0.52 = 0.4940$

\item \textbf{STM2 temporal fusion} ($q = 0.8409$):
$\piN = 0.5843$, $\piA = 0.4157$

\item \textbf{Bayesian update:}
$P(N)_{\text{post}} = \colorbox{lightyellow}{\,$0.5901$\,}$

\item \textbf{Decision:} $0.30 < 0.5901 < 0.70$ $\to$ \inconclusive. Carry forward, shrink window.
\end{enumerate}

\subsection{Round 2 \texorpdfstring{$(W = 4.0$}{(W=4.0}\,s) --- \texorpdfstring{$\{M_1, M_3, M_4\}$}{\{M1,M3,M4\}}, Wtd.\ Sum}

Face becomes available. $\hat{w} = (0.71, 0.18, 0.12)$.

\begin{enumerate}[itemsep=3pt, parsep=1pt, label=\textbf{\arabic*.}]
\item \textbf{Entry state:} $\pn = 0.5901$, $\pa = 0.4099$

\item \textbf{Per-modality scores and fusion:}
\begin{center}
\small
\begin{tabular}{@{}l cc@{}}
\toprule
& $P_j^N$ & $P_j^{\lnot N}$ \\
\midrule
$M_1$ (face) & 0.91 & 0.09 \\
$M_3$ (voice) & 0.58 & 0.42 \\
$M_4$ (keystroke) & 0.52 & 0.48 \\
\bottomrule
\end{tabular}
\end{center}
$L_N = 0.71 \times 0.91 + 0.18 \times 0.58 + 0.12 \times 0.52 = 0.8129$\\
$L_{\lnot N} = 0.71 \times 0.09 + 0.18 \times 0.42 + 0.12 \times 0.48 = 0.1971$

\item \textbf{STM2 temporal fusion} ($q = e^{-\frac{\ln 2}{20} \times 4.0} = 0.8706$):
$\piN = 0.8706 \times 0.5901 = 0.5137$, \quad $\piA = 0.1294 \times 0.5901 + 0.4099 = 0.4863$

\item \textbf{Bayesian update:}
$P(N)_{\text{post}} = \colorbox{lightgreen}{\,$0.8133$\,}$

\item \textbf{Decision:} $0.8133 \geq 0.70$ $\to$ \recover.
\end{enumerate}

\medskip
\noindent\textit{Outcome:} Return to P1 with $\pn = 0.8133$. Penalty $= 2\delta = 2.0$\,s. If re-entered immediately: $W_{\text{eff}} = \max(1.0,\; 5.0 - 2.0) = 3.0$\,s.

\section{Scenario C: Gradual Alert}
\label{sec:scenario_c}

An attacker with limited biometric capability. Each round uses a different subset as the attacker tries different sensors. All produce weak scores.

\subsection{Round 1 \texorpdfstring{$(W = 5.0$}{(W=5.0}\,s) --- \texorpdfstring{$\{M_4\}$}{\{M4\}} single, Product rule}

\begin{enumerate}[itemsep=3pt, parsep=1pt, label=\textbf{\arabic*.}]
\item \textbf{Entry state:} $\pn = 0.6948$, $\pa = 0.3052$

\item \textbf{Fusion:} Single modality, Product rule.
$L_N = 0.40$, $L_{\lnot N} = 0.60$

\item \textbf{STM2 temporal fusion} ($q = 0.8409$):
$\piN = 0.5843$, $\piA = 0.4157$

\item \textbf{Bayesian update:}
$P(N)_{\text{post}} = \colorbox{lightyellow}{\,$0.4837$\,}$

\item \textbf{Decision:} $0.30 < 0.4837 < 0.70$ $\to$ \inconclusive.
\end{enumerate}

\subsection{Round 2 \texorpdfstring{$(W = 4.0$}{(W=4.0}\,s) --- \texorpdfstring{$\{M_3, M_5\}$}{\{M3,M5\}}, Wtd.\ Sum}

\begin{enumerate}[itemsep=3pt, parsep=1pt, label=\textbf{\arabic*.}]
\item \textbf{Entry state:} $\pn = 0.4837$, $\pa = 0.5163$

\item \textbf{Fusion} ($\hat{w} = [0.65, 0.35]$): $M_3$: $(0.42, 0.58)$, $M_5$: $(0.38, 0.62)$.\\
$L_N = 0.4060$, $L_{\lnot N} = 0.5940$

\item \textbf{STM2 temporal fusion} ($q = 0.8706$):
$\piN = 0.8706 \times 0.4837 = 0.4211$, \quad $\piA = 0.1294 \times 0.4837 + 0.5163 = 0.5789$

\item \textbf{Bayesian update:}
$P(N)_{\text{post}} = \colorbox{lightyellow}{\,$0.3321$\,}$

\item \textbf{Decision:} $0.30 < 0.3321 < 0.70$ $\to$ \inconclusive.
\end{enumerate}

\subsection{Round 3 \texorpdfstring{$(W = 3.0$}{(W=3.0}\,s) --- \texorpdfstring{$\{M_3, M_4, M_5\}$}{\{M3,M4,M5\}}, Wtd.\ Sum}

\begin{enumerate}[itemsep=3pt, parsep=1pt, label=\textbf{\arabic*.}]
\item \textbf{Entry state:} $\pn = 0.3321$, $\pa = 0.6679$

\item \textbf{Fusion} ($\hat{w} = [0.46, 0.30, 0.24]$): $M_3$: $(0.42, 0.58)$, $M_4$: $(0.38, 0.62)$, $M_5$: $(0.35, 0.65)$.\\
$L_N = 0.3912$, $L_{\lnot N} = 0.6088$

\item \textbf{STM2 temporal fusion} ($q = e^{-\frac{\ln 2}{20} \times 3.0} = 0.9013$):
$\piN = 0.9013 \times 0.3321 = 0.2993$, \quad $\piA = 0.0987 \times 0.3321 + 0.6679 = 0.7007$

\item \textbf{Bayesian update:}
$P(N)_{\text{post}} = \colorbox{lightred}{\,$0.2153$\,}$

\item \textbf{Decision:} $0.2153 \leq \Talert = 0.30$ $\to$ \alert.
\end{enumerate}

\medskip
\noindent\textit{Outcome:} Session locked. Total P2 duration: $5.0 + 4.0 + 3.0 = 12.0$\,s.

$\pa$ progression: $0.3052 \to 0.5163 \to 0.6679 \to 0.7847$. Compounding pressure forces the outcome.

\section{Scenario D: No Observations}
\label{sec:scenario_d}

\begin{enumerate}[itemsep=3pt, parsep=1pt, label=\textbf{\arabic*.}]
\item \textbf{Entry state:} $\pn = 0.6948$, $\pa = 0.3052$. Window $= 5.0$\,s.
\item \textbf{System waits} for observations. None arrive.
\item \textbf{Decision:} No observations $\to$ \alert{} (immediate, no further rounds).
\end{enumerate}

\medskip
This prevents an attacker from stalling by avoiding sensors entirely.

\section{Scenario E: Max-Rounds Exhaustion}
\label{sec:scenario_e}

Attacker produces perfectly neutral evidence ($L_N = L_{\lnot N} = 0.04$) every round. Equal likelihoods mean the Bayesian update preserves the prior ($P(N)_{\text{post}} = \piN$), so $\pn$ decreases purely from STM2 decay.

\begin{table}[H]
\centering
\caption{Scenario E: $P(N)$ trajectory with equal evidence ($L_N = L_{\lnot N} = 0.04$).}
\label{tab:scenario_e}
\small
\begin{tabular}{@{}c c c c l@{}}
\toprule
\textbf{Round} & $\boldsymbol{W}$ & $\boldsymbol{q}$ & $\boldsymbol{P(N)_{\text{post}}}$ & \textbf{Decision} \\
\midrule
\rowcolor{lightyellow}
1 & 5.0\,s & 0.8409 & 0.5843 & Inconclusive \\
\rowcolor{lightyellow}
2 & 4.0\,s & 0.8706 & 0.5086 & Inconclusive \\
\rowcolor{lightyellow}
3 & 3.0\,s & 0.9013 & 0.4584 & Inconclusive \\
\rowcolor{lightyellow}
4 & 2.0\,s & 0.9330 & 0.4277 & Inconclusive \\
\rowcolor{lightyellow}
5 & 1.0\,s & 0.9659 & 0.4131 & Inconclusive \\
\midrule
\multicolumn{4}{@{}l}{Max rounds exhausted} & \textcolor{cRed}{\textbf{ALERT}} \\
\bottomrule
\end{tabular}
\end{table}

Total P2 duration: $5.0 + 4.0 + 3.0 + 2.0 + 1.0 = 15.0$\,s. Even with neutral evidence, STM2 decay erodes $\pn$ monotonically. Shorter windows reduce per-round decay ($q$ closer to 1), but the cumulative effect is decisive.

\section{\texorpdfstring{Scenario F: P2 $\to$ P1 $\to$ P2 Re-Entry}{Scenario F: P2 -> P1 -> P2 Re-Entry}}
\label{sec:scenario_f}

Demonstrates the full lifecycle: recovery, return to P1, penalty decay, and a second P2 entry with restored window.

\subsection{First P2: Immediate Recovery (same as Scenario A)}

\begin{enumerate}[itemsep=3pt, parsep=1pt, label=\textbf{\arabic*.}]
\item Entry: $\pn = 0.6948$, $\pa = 0.3052$. Penalty $= 0$. Window $= 5.0$\,s.
\item R1 with $\{M_1, M_2, M_3\}$ Wtd.\ Geom.\ $\to$ $P(N)_{\text{post}} = 0.9519$.
\item Decision: $0.9519 \geq 0.70$ $\to$ \recover. Penalty $\to 1.0$\,s.
\end{enumerate}

\subsection{Return to P1 (20\,s in Normal)}

\begin{enumerate}[itemsep=3pt, parsep=1pt, label=\textbf{\arabic*.}]
\item Re-enter P1: $\pn = 0.9519$, $\ps = 0.0481$. Penalty $= 1.0$\,s.
\item Penalty decay over 20\,s: $\max(0,\; 1.0 - 0.1 \times 20) = 0.0$. Full window restored.
\item P1 decay (no obs): $p = e^{-\frac{\ln 2}{10} \times 20} = 0.2500$.\\
$\piN = 0.2500 \times 0.9519 = 0.2380$, \quad $\piS = 0.7500 \times 0.9519 + 0.0481 = 0.7620$

Without new observations, $\pn$ decays to $0.2380$. A weak observation triggers P2 again.
\end{enumerate}

\subsection{Second P2 Entry --- \texorpdfstring{$\{M_1, M_2\}$}{\{M1,M2\}}, Wtd.\ Geom.}

\begin{enumerate}[itemsep=3pt, parsep=1pt, label=\textbf{\arabic*.}]
\item Entry: $\pn = 0.55$ (after weak P1 obs), $\pa = 0.45$. Penalty $= 0$. Window $= 5.0$\,s (restored).

\item \textbf{Fusion} ($\hat{w} = [0.43, 0.57]$): $M_1$: $(0.91, 0.09)$, $M_2$: $(0.94, 0.06)$.\\
$L_N = 0.91^{0.43} \times 0.94^{0.57} = 0.9270$, \quad $L_{\lnot N} = 0.09^{0.43} \times 0.06^{0.57} = 0.0714$

\item \textbf{STM2 temporal fusion} ($q = 0.8409$):
$\piN = 0.8409 \times 0.55 = 0.4625$, \quad $\piA = 0.1591 \times 0.55 + 0.45 = 0.5375$

\item \textbf{Bayesian update:}
$P(N)_{\text{post}} = \colorbox{lightgreen}{\,$0.9178$\,}$

\item \textbf{Decision:} $0.9178 \geq 0.70$ $\to$ \recover. Penalty $\to 1.0$\,s.
\end{enumerate}

\medskip
\noindent\textit{Key insight:} The penalty fully decayed during 20\,s of Normal, so the second P2 uses the full 5.0\,s window. VIGIL is fair to genuine users who maintain stable behavior between P2 episodes.

\section{Adaptive Window Session Timeline}
\label{sec:adaptive_window_supp}

Table~\ref{tab:adaptive_window_supp} traces penalty accumulation and decay across multiple P2 entries.

\begin{table}[H]
\centering
\caption{Adaptive window session ($W_{base}\!=\!5.0$\,s, $\delta\!=\!1.0$\,s, $\lambda\!=\!0.1$\,s$^{-1}$, $W_{\min}\!=\!1.0$\,s).}
\label{tab:adaptive_window_supp}
\small
\begin{tabular}{@{}r l c c@{}}
\toprule
$\boldsymbol{t}$ & \textbf{Event} & \textbf{Penalty} & $\boldsymbol{W_{\text{eff}}}$ \\
\midrule
\rowcolor{red!8}
$0$\,s & 1st P2 entry & $0.0$ & $5.0$\,s \\
\rowcolor{red!14}
$5$\,s & 2nd P2 entry (immediate) & $1.0$ & $4.0$\,s \\
\rowcolor{red!20}
$9$\,s & 3rd P2 entry (immediate) & $2.0$ & $3.0$\,s \\
\rowcolor{green!14}
$12$\,s & Recovers to Normal & $3.0$ & $2.0$\,s \\
\rowcolor{green!14}
$22$\,s & 10\,s in Normal & $2.0$ & $3.0$\,s \\
\rowcolor{red!14}
$32$\,s & 4th P2 entry & $1.0$ & $4.0$\,s \\
\rowcolor{green!14}
$36$\,s & Recovers to Normal & $2.0$ & $3.0$\,s \\
\rowcolor{green!14}
$56$\,s & 20\,s in Normal (full reset) & $0.0$ & $5.0$\,s \\
\bottomrule
\end{tabular}
\end{table}

\colorbox{red!14}{Red rows}: P2 entries (penalty $+\delta$). \colorbox{green!14}{Green rows}: Normal periods (penalty decays at $\lambda$). A persistent threat faces shrinking windows; a genuine user's penalty resets.

\section{Summary}
\label{sec:summary}

\begin{table}[H]
\centering
\caption{Summary of all P2 verification scenarios.}
\label{tab:summary}
\small
\begin{tabular}{@{}l c l c p{3.8cm}@{}}
\toprule
\textbf{Scenario} & \textbf{Rnds} & \textbf{Outcome} & \textbf{Time} & \textbf{Key Feature} \\
\midrule
\rowcolor{lightgreen}
A: Immediate Recovery & 1 & \recover & 5.0\,s & Strong $\{M_1,M_2,M_3\}$ \\
\rowcolor{lightgreen}
B: Multi-Round & 2 & \recover & 9.0\,s & $\{M_3,M_5\}$ borderline $\to$ $\{M_1,M_3,M_4\}$ strong \\
\rowcolor{lightred}
C: Gradual Alert & 3 & \alert & 12.0\,s & Weak across 3 subsets \\
\rowcolor{lightred}
D: No Observations & 0 & \alert & 5.0\,s & Stalling prevention \\
\rowcolor{lightred}
E: Max-Rounds & 5 & \alert & 15.0\,s & Neutral $L_N\!=\!L_{\lnot N}$, decay only \\
\rowcolor{lightgreen}
F: P2$\to$P1$\to$P2 & 1+1 & \recover & --- & Penalty decay restores window \\
\bottomrule
\end{tabular}
\end{table}

These six scenarios demonstrate all terminal paths through VIGIL:
\begin{enumerate}[itemsep=2pt, parsep=0pt]
    \item \textit{Recovery via strong evidence} (A) --- $P(N)_{\text{post}} \geq \Tback$ in R1.
    \item \textit{Recovery via accumulated evidence} (B) --- inconclusive then strong, different subsets.
    \item \textit{Alert via weak evidence} (C) --- attacker tries different sensors, $\pa$ compounds.
    \item \textit{Alert via sensor avoidance} (D) --- no observations $\to$ immediate lockout.
    \item \textit{Alert via exhaustion} (E) --- all rounds inconclusive, max rounds exceeded.
    \item \textit{Re-entry with restoration} (F) --- penalty decays, window resets for genuine users.
\end{enumerate}

\end{document}